\documentclass[%
 reprint,
 superscriptaddress,
 amsmath,amssymb,
 aps,
 pra,
 longbibliography,
lengthcheck,%
]{revtex4-1}

\usepackage{graphicx}
\usepackage{dcolumn}
\usepackage{bm}
\usepackage{hyperref}
\usepackage{color}
\usepackage{ulem}
\usepackage{natbib}
\usepackage{xr}

\renewcommand{\Re}[0]{{\rm Re}}
\renewcommand{\Im}[0]{{\rm Im}}

\begin{document}

\preprint{APS/123-QED}

\title{
Nonlinear response theory for orbital photocurrent in semiconductors
}

\author{Kakeru Tanaka}
\affiliation{
Department of Physics, Institute of Science Tokyo, Meguro, Tokyo, 152-8551, JAPAN 
}

\author{Hiroaki Ishizuka}
\affiliation{
Department of Physics, Institute of Science Tokyo, Meguro, Tokyo, 152-8551, JAPAN 
}

\date{\today}

\begin{abstract}
Recent theoretical studies on the nonlinear response of spin and orbital degrees of freedom have discovered spin and orbital analogs of the photocurrent, with potential for characterizing topological materials and for applications.
In this paper, we develop a general theory for calculating spin and orbital currents in semiconductors and study the properties of optical responses in the Bernevig-Hughes-Zhang and Luttinger models, where nonlinear orbital responses and a topological phase transition occur.
We study the evolution of optical responses at the topological phase transition and how they manifest.
In addition, we find that the relaxation time dependence of the orbital conductivity is somewhat distinct from that of the photocurrent.
The theory is straightforwardly applicable to complex models of real materials, allowing quantitative predictions of the nonlinear responses of orbital and spin.
\end{abstract}

\maketitle

\section{Introduction}

The second-order response to electromagnetic waves and light has been extensively studied due to its importance in physics and applications.
One such phenomenon is the photocurrent, the dc current induced by incident light~\cite{Nelson2003a,Wurfel2005a}, as in semiconductor pn junction solar cells.
In addition to pn junction solar cells, a photocurrent has also been known to occur in bulk materials, which is known as the bulk photovoltaic effect~\cite{vBaltz1981a,Belinicher1982a,Sturman1992a,Sipe2000a,Young2012a,Cook2017a,Tokura2018a}.
This phenomenon has also been studied for a long time.
However, the phenomenon has recently attracted renewed interest from the viewpoints of quantum geometry and topological materials, such as the Berry phase~\cite{Moore2010a,Sodemann2015a,Ishizuka2016a,Ishizuka2017b,Morimoto2018a} and the Riemannian metric~\cite{Ahn2022a}.
These studies demonstrate the rich physics underlying the photocurrent and its potential for studying the topological properties of solids.

In addition to the photocurrent, recent theoretical works have pointed out the possibility of similar phenomena in spin and orbital currents.
These phenomena include spin and orbital currents carried by excited electrons similar to the bulk photovoltaic effect~\cite{Young2013a,Davydova2022a}, as well as those by other excitations such as magnons~\cite{Proskurin2018a,Ishizuka2019a,Ishizuka2022a}, spinons~\cite{Ishizuka2019b}, and phonons~\cite{Ishizuka2024a,Ishizuka2025a};
some of these phenomena have also been shown to be related to the Berry connection of excitations~\cite{Fujiwara2023a,Ishizuka2024a}.
For the spin current, there has been a wide range of theoretical studies that provide general formulas, such as nonlinear response theory~\cite{Young2013a,Ishizuka2019a,Xu2021a}, which provide a generic yet convenient framework for studying exotic nonlinear responses. These formulas are becoming increasingly important, given that spin photocurrents have recently been theoretically explored in a wide variety of materials~\cite{Wang2024a,Huang2024,Zhu2024,Zhang2025,Yang2025,Movafagh2025}.

In addition to these studies, recent studies find that a phenomenon similar to the spin photocurrent also occurs in centrosymmetric materials.
For instance, a recent study finds that an orbital response related to electrical polarization can occur in a centrosymmetric material~\cite{Davydova2022a}.
Such an effect, if confirmed experimentally, may further extend the study of the novel optical response in bulk materials.
However, a generic formula for studying the orbital photocurrent in a model relevant to actual materials has not yet been given.

In this study, we study a generic formula for the orbital and spin photocurrent based on the nonlinear response theory.
Following the prescription by Sipe {\it et al.}~\cite{Sipe2000a}, we derive a generic formula for the second-order orbital and spin photocurrents.
For the dc current, we show that the formula consists of the shift and injection current contributions, each of which shows different relaxation time dependence.
Using this theory, we study the orbital and spin currents in the Bernivig-Hughes-Zhang (BHZ)\cite{Bernevig2006a} and Luttinger models~\cite{Luttinger1956}, the models in which a topological phase transition occurs.
For the response to linearly-polarized light in the BHZ model, we reproduce a result presented in a previous work~\cite{Davydova2022a}.
Additionally, we find that an orbital current analogous to the injection current occurs due to the circularly polarized light.
In addition, we show that the spin current occurs in the Luttinger model in the presence of Rashba spin-orbit interaction.

\section{Nonlinear response theory}\label{sec:NLRT}

\subsection{Nonlinear conductivity}
We consider a non-interacting electron system, whose Hamiltonian $\mathcal{H}_0$ coupled to the electric field $\bm{E}(t)$.
The Hamiltonian reads
\begin{align}
H(t)&=\int\psi^\dagger(\bm{x},t)[\mathcal{H}_0-e\bm{x}\cdot\bm{E}(t)]\psi(\bm{x},t)d\bm{x}
\end{align}
where $e$ ($<0$) is the charge of electrons, $\bm{x}$ is the position operator, $\psi(\bm{x},t)$ and $\psi^\dagger(\bm{x},t)$ are the field operator for electrons at position $\bm{x}$ and time $t$ and its conjugate, respectively. 

Using the eigenstate wavefunctions of ${\cal H}_0$, $\psi_{n,\bm{k}}(\bm{x},t)$, the field operator reads
\begin{align}
\psi(\bm{x},t)=\sum_n\int d\bm{k}\,c_n(\bm{k})\psi_{n,\bm{k}}(\bm{x},t) \label{eq:phi}
\end{align}
where $n$ is the band index, $\bm{k}$ is the wave vector and $c_n(\bm{k})$ are the annihilation operators, and $\psi_{n,\bm{k}}(\bm{x},t)$ are the eigenstates of $\mathcal{H}_0$ for the $n$-th band with momentum $\bm{k}$.
These eigenstates satisfy
\begin{align}
\mathcal{H}_0\psi_{n,\bm{k}}(\bm{x},t)=\epsilon_n(\bm{k})\psi_{n,\bm{k}}(\bm{x},t),
\end{align}
where $\epsilon_n(\bm{k})$ is the eigenenergy of $n$-th band state with the wave vector $\bm k$.
The integral in Eq.~\eqref{eq:phi} is over the first Brillouin zone.

The spin and orbital current operators are written as
\begin{align}
\bm{J}(t)=\sum_{n,m}\int\frac{d\bm{k}}{\Omega}\frac{1}{2}\{\hat{O},\hat{\bm{v}}\}_{nm}c_n^\dagger(\bm{k})c_m(\bm{k}),\label{soo}
\end{align}
where $\Omega$ is the volume of the lattice, $\hat{\bm{v}}=\frac1\hbar\partial_{\bm{k}}\mathcal{H}_0$ is the velocity operator, and $\hat{O}$ is the spin or orbital operator~\cite{Sinova2015a}. Each element of this operator reads $v^a_{nm}(\bm{k})=\langle n,\bm{k}|\hat{v}^a|m,\bm{k}\rangle, O_{nm}(\bm{k})=\langle n,\bm{k}|\hat{O}|m,\bm{k}\rangle$ where $a,b,c=x,y,z$ are the Cartesian coordinates.

To study the linear and nonlinear responses, we expand the density matrix,
\begin{align}
\rho_{mn}(\bm{k})=\frac{\Omega}{8\pi^3}\langle c_n^\dagger(\bm{k})c_m(\bm{k})\rangle,
\end{align}
to $\rho_{mn}=\rho_{mn}^{(0)}+\rho_{mn}^{(1)}+\rho_{mn}^{(2)}+\dots$, where $\rho_{mn}^{(n)}$ is the $n$th order term with respect to $\bm E$~\cite{Sipe2000a}.
From the Heisenberg equation of motion for the density operator $a_n^\dagger a_m$, we obtain the differential equation for $\rho_{mn}$,
\begin{align}
&\frac{\partial \rho_{mn}}{\partial t}+\frac{i}{\hbar}\epsilon_{mn}\rho_{mn}\nonumber\\
&=-\frac{eE^b_\beta e^{-i\omega_\beta t}}{\hbar}\rho_{mn;b}+\frac{ieE^b_\beta e^{-i\omega_\beta t}}{\hbar}
\sum_p(r^b_{mp}\rho_{pm}-\rho_{mp}r^b_{pn}).
\label{eq:rho}
\end{align}
Here, we assumed the $b$ component of electric field reads $E^b(t)=\sum_\beta E^b_\beta e^{-i\omega_\beta t}$.

We first consider the first-order response.
To this end, we assume that the system is an insulator and the initial state is the ground state, i.e., $\rho^{(0)}_{mn}=f_n\delta_{mn}$, where $f_n=1$ ($f_n=0$) for the valence (conduction) bands. 
By using Eq.~\eqref{eq:rho}, the equation of motion for $\rho_{mn}^{(1)}$ reads
\begin{align}
\frac{\partial \rho^{(1)}_{mn}}{\partial t}+\frac{i}{\hbar}\epsilon_{mn}\rho^{(1)}_{mn}
=-\frac{ieE^b_\beta e^{-i\omega_\beta t}}{\hbar}f_{nm}r^b_{mn}.
\end{align}
The solution of this equation is
\begin{align}
\rho^{(1)}_{mn}=\mathcal{B}^b_{mn}E^b_\beta e^{-i\omega_\beta t},
\label{eq:1st_rho}
\end{align}
where,
\begin{align}
\mathcal{B}^b_{mn}&=\frac{ef_{nm}r^b_{mn}}{\epsilon_{mn}-\hbar\omega_\beta-i\eta},\\
\bm{\xi}_{nm}&=i\langle n,\bm{k}|\partial_{\bm{k}}|m,\bm{k}\rangle,\\
r^a_{nm}&=\left\{
\begin{aligned}
&\xi^a_{nm}&&n\neq m\\
&0&&n=m,
\end{aligned}
\right.
\end{align}
Here, $f_{nm}(\bm k)=f_n(\bm k)-f_m(\bm k)$ is the difference of the fermi distribution function $f_n(\bm k)=1/(e^{\beta \epsilon_n(\bm k)}+1)$ and $\epsilon_{nm}(\bm k)=\epsilon_n(\bm k)-\epsilon_m(\bm k)$. 

Using Eq.~\eqref{eq:1st_rho}, the first-order response reads,
\begin{align}
&\sigma^{(1)}_{ab}(\omega_\beta)\nonumber\\
&=-i\hbar\int_{\bm{k}}\sum_{n\mu m}\frac{f_{nm}}{\epsilon_{nm}}\frac{\frac{1}{2}(O_{n\mu}v^a_{\mu m}+v^a_{n\mu}O_{\mu m})(ev^b_{mn})}{\hbar\omega+\epsilon_{nm}+i\eta},\label{eq:1st}
\end{align}
where $\int_{\bm{k}}=\int \frac{d\bm{k}}{8\pi^3}$.
This formula is identical to the Kubo formula for spin or orbital current. 

Next, we consider the second-order response.
Similarly to the first-order response, the equation of motion for $\rho_{mn}^{(2)}$ is,
\begin{align}
\frac{\partial \rho^{(2)}_{mn}}{\partial t}+\frac{i}{\hbar}\epsilon_{mn}\rho_{mn}^{(2)}
&=-\frac{eE^b_\beta E^c_\gamma e^{-i\omega_\Sigma t}}{\hbar}\mathcal{B}^b_{mn;c}\nonumber\\
\quad+&\frac{ieE^b_\beta E^c_{\gamma} e^{-i\omega_\Sigma t}}{\hbar}
\sum_p(r^b_{mp}\mathcal{B}_{pm}-\mathcal{B}_{mp}r^b_{pn}).
\end{align}
Solving this equation yields the following result for the second-order term,
\begin{align}
&\rho^{(2)}_{mn}(\bm k)\nonumber\\
&=\frac{e}{i(\epsilon_{mn}-\hbar\omega_{\Sigma}-i\eta)}\left[\mathcal{B}_{mn;c}^b-i\sum_p(r^c_{mp}\mathcal{B}^b_{pn}
-\mathcal{B}^b_{mp}r^c_{pn})\right]\nonumber\\
&\ \ \ \times E^b_\beta E^c_\gamma e^{-i\omega_{\Sigma}t}.\label{eq:rho_E2}
\end{align}
Here, $O_{nm;c}$ is the generalized derivative of $O_{nm}$ defined by 
\begin{align}
O^a_{nm;c}(\bm{k})\equiv\frac{\partial O^a_{nm}(\bm{k})}{\partial k^c}-i[\xi^c_{nn}(\bm{k})-\xi^c_{mm}(\bm{k})]O^a_{nm}(\bm{k}).
\end{align}
Using Eq.~\eqref{eq:rho_E2}, the second-order response in spin and orbital degrees of freedom reads
\begin{widetext}
\begin{align}
&\sigma_{abc}^{(2)}(\omega_\Sigma;\omega_\beta,\omega_\gamma)\nonumber\\
&=\frac{ie^2}{2}\int_{\bm{k}}\sum_{n m}f_{nm}s^a_{nm}
\left[\left(\frac{r_{mn;c}^b}{\rho_\gamma\epsilon_{mn}}-\frac{\hbar r_{mn}^b\Delta_{mn}^c}{\rho_\gamma^2\epsilon_{mn}^2}\right)
+(bc\beta\gamma\Leftrightarrow cb\gamma\beta)\right]\frac{1}{\epsilon_{mn}-\hbar\omega_\Sigma-2i\eta}\nonumber\\
&\quad+\frac{ie^2}{2}\int_{\bm{k}}\sum_{nm}f_{nm} 
\left\{\left[r_{mn}^b\left(\frac{\rho_\beta s^a_{nm;c}}{\rho_\gamma\epsilon_{mn}}
+\frac{\hbar\rho_\beta^2s_{nm}^a\Delta_{mn}^c}{\rho_\gamma^2\epsilon_{mn}^2}\right)\frac{1}{\epsilon_{mn}-\hbar\omega_\beta-i\eta}\right]
+(bc\beta\gamma\Leftrightarrow cb\gamma\beta)\right\}\nonumber\\
&\quad+\frac{e^2}{2}\int_{\bm{k}}\sum_{nmp}\left\{\left[s^a_{nm}\frac{r_{mp}^br_{pn}^c}{\epsilon_{mp}-\rho_\beta\epsilon_{mn}}\left(\frac{\rho_\beta f_{pm}}{\epsilon_{mp}-\hbar\omega_\beta-i\eta}
+\frac{\rho_\gamma f_{np}}{\epsilon_{pn}-\hbar\omega_\gamma-i\eta}-\frac{f_{nm}}{\epsilon_{mn}-\hbar\omega_\Sigma-2i\eta}\right)\right]+(bc\beta\gamma\Leftrightarrow cb\gamma\beta)\right\},\label{2nd}
\end{align}
\end{widetext}
where $s^a_{nm}=\sum_\mu\frac{1}{2}(O_{n\mu}v^a_{\mu m}+v^a_{n\mu}O_{\mu m})$, $\omega_\Sigma=\omega_\beta+\omega_\gamma, \rho_\beta=\omega_\beta/\omega_\Sigma$, and $\rho_\gamma=\omega_\gamma/\omega_\Sigma$.

In the case of the photocurrent, i.e., $\omega_\Sigma=0$, it is known that the formula for the conductivity can be separated into two contributions, the shift and injection currents~\cite{Sturman1992a,Sipe2000a}.
Applying this notion, in the following, we separate the second-order dc response tensor in Eq.~\eqref{2nd} 
into two contributions, 
$$\sigma^{(2)}_{abc}(0;\omega;-\omega)=\sigma^{(2A)}_{abc}(0;\omega;-\omega)+\sigma^{(2B)}_{abc}(0;\omega;-\omega).$$
where
\begin{align}
    &\sigma_{abc}^{(2A)}(0;\omega;-\omega)=\nonumber\\
    &\frac{ie^2}{2}\int_{\bm{k}}\sum_{nm}f_{nm}r_{mn}^b\left(-\frac{s_{nm;c}^a}{\epsilon_{mn}}
+\frac{\hbar s^a_{nm}\Delta_{mn}^c}{\epsilon_{mn}^2}\right)\frac{1}{\epsilon_{mn}-\hbar\omega-i\eta}\nonumber\\
&\qquad+(bc\beta\gamma\Leftrightarrow cb\gamma\beta)\nonumber\\
&-\frac{e^2}{2}\int_{\bm{k}}\sum_{nmp}\frac{f_{nm}s_{nm}^ar_{mp}^br_{pn}^c}{\epsilon_{mn}}\left(\frac{f_{pm}}{\epsilon_{mp}-\hbar\omega-i\eta}
-\frac{f_{np}}{\epsilon_{pn}+\hbar\omega-i\eta}\right)\nonumber\\
&\qquad+(bc\beta\gamma\Leftrightarrow cb\gamma\beta),
\label{eq:shift_pre}
\end{align}
and
\begin{align}
    &\sigma_{abc}^{(2B)}(0;\omega;-\omega)
    =\nonumber\\
    &\frac{e^2}{2}\int_{\bm{k}}\sum_{nmp}\left[\frac{s_{nm}^ar_{mp}^br_{pn}^c}{\epsilon_{mp}-\rho_\beta\epsilon_{mn}}\right.\left(\frac{\rho_\beta f_{pm}}{\epsilon_{mp}-\hbar\omega_\beta-i\eta}
+\frac{\rho_\gamma f_{np}}{\epsilon_{pn}-\hbar\omega_\gamma-i\eta}\right)\nonumber\\
    &\quad\left.+(bc\beta\gamma\Leftrightarrow cb\gamma\beta)\right].
\label{eq:inj_pre}
\end{align}
Here, we defined $\sigma^{(2A)}_{abc}(0;\omega;-\omega)$ as a current independent of the relaxation time in the clean limit, and $\sigma^{(2B)}_{abc}(0;\omega;-\omega)$ as a current proportional to the relaxation time.
They correspond to the shift current and the injection current, respectively.
In the above, we focus on the dc limit, which corresponds to substituting $\omega_\beta=\omega$ and $\omega_\gamma=-\omega$.


\subsection{Shift current}
Let us first look at $\sigma^{(2A)}_{abc}(0;\omega;-\omega)$.
By using the shift vector,
\begin{widetext}
\begin{align}
\sigma_{abc}^{(2A)}(0;\omega;-\omega)
&=\frac{ie^2}{2}\int_{\bm{k}}\sum_{nm}f_{nm}|r_{mn}^b|e^{-\phi^b_{nm}}
\left(-\frac{[\partial_c(|O_{n\mu}||v^a_{\mu m}|)-i(R^{O,c}_{n\mu}+R^{a,c}_{\mu m})|O_{n\mu}||v^a_{\mu m}|]e^{-i\phi^O_{n\mu}-i\phi^a_{\mu m}}}{\epsilon_{mn}}\right.\nonumber\\
&\ \ \ \ \ \ \ \left.-\frac{[\partial_c(|v^a_{n\mu}||O_{\mu m}|)-i(R^{a,c}_{n\mu}+R^{O,c}_{\mu m})|v^a_{n\mu}||O_{\mu m}|]e^{-i\phi^a_{n\mu}-i\phi^O_{\mu m}}}{\epsilon_{mn}}\right.\nonumber\\
&\ \ \ \ \ \ \ \left.
+\frac{\hbar(|O_{n\mu}||v^a_{\mu m}|e^{-i\phi^O_{n\mu}-i\phi^a_{\mu m}}+|v^a_{n\mu}||O_{\mu m}|e^{-i\phi^a_{n\mu}-i\phi^O_{\mu m}})\Delta_{mn}^c}{\epsilon_{mn}^2}\right)\frac{1}{\epsilon_{mn}-\hbar\omega-i\eta}\nonumber\\
&\ \ \ \ \ +(bc\beta\gamma\Leftrightarrow cb\gamma\beta)\nonumber\\
&-\frac{e^2}{2}\int_{\bm{k}}\sum_{nmp}\left[\frac{f_{nm}(|O_{n\mu}||v^a_{\mu m}|e^{-i\phi^O_{n\mu}-i\phi^a_{\mu m}}+|v^a_{n\mu}||O_{\mu m}|e^{-i\phi^a_{n\mu}-i\phi^O_{\mu m}})|r_{mp}^b||r_{pn}^c|e^{-i\phi^b_{mp}-i\phi^c_{pn}}}{\epsilon_{mn}}\right.\nonumber\\
&\ \ \ \ \ \left.\times\left(\frac{f_{pm}}{\epsilon_{mp}-\hbar\omega-i\eta}
-\frac{f_{np}}{\epsilon_{pn}+\hbar\omega-i\eta}\right)+(bc\beta\gamma\Leftrightarrow cb\gamma\beta)\right],\label{eq:shift}
\end{align}
\end{widetext}
where,
\begin{align}
v^a_{nm}&=|v^*_{nm}|e^{-i\phi^*_{nm}},\label{phi1}\\
O_{nm}&=|O_{nm}|e^{-i\phi^O_{nm}},\label{phi2}\\
R^{a,b}_{nm}&=\frac{\partial\phi^a_{nm}}{\partial k^b}+\xi^b_{nn}-\xi^b_{mm}.
\end{align}


To understand how symmetry affects $\sigma_{abc}^{(2A)}$, we first consider systems with time-reversal symmetry. 
In the presence of the time-reversal symmetry, the eigenstates for $-\bm k$ are given by $|u_{n,-\bm{k}}\rangle=\hat{T}|u_{n,\bm{k}}\rangle$, where $\hat{T}$ is a time-reversal operator.
Using this basis, one can show the following relations,
\begin{align}
\epsilon_{nm}(-\bm{k})&=\epsilon_{nm}(\bm{k}),\label{tim1}\\
\bm{r}_{nm}(-\bm{k})&=\bm{r}_{mn}(\bm{k}),\\
\bm{v}_{nm}(-\bm{k})&=-\bm{v}_{mn}(\bm{k}),\\
O_{nm}(-\bm{k})&=(\hat{T}^{-1}\hat{O}\hat{T})_{nm}(\bm{k}).\label{tim4}
\end{align}
For example, $O_{nm}(-\bm{k})=-O_{mn}(\bm{k})$ when $\hat{O}$ is a spin operator.
Using the relations in Eqs.~\eqref{tim1}-\eqref{tim4}, the conductivity for linearly polarized light $\sigma_{abb}^{(2A)}(0;\omega;-\omega)$ reads,
\begin{widetext}
\begin{align}
\sigma_{abb}^{(2A)}(0;\omega;-\omega)
&=\frac{i e^2}{2}\int_{\bm{k}}\sum_{nm}f_{nm}r_{mn}^b\left(-\frac{s_{nm;b}^a(\bm{k})}{\epsilon_{mn}}
+\frac{\hbar s^a_{nm}(\bm{k})\Delta_{mn}^b}{\epsilon_{mn}^2}\right)\frac{1}{\epsilon_{mn}-\hbar\omega-i\eta}\nonumber\\
&\quad-\frac{i e^2}{2}\int_{\bm{k}}\sum_{nm}f_{nm}r_{mn}^b\left(-\frac{s_{nm;b}^a(\bm{-k})}{\epsilon_{mn}}
-\frac{\hbar s^a_{nm}(\bm{-k})\Delta_{mn}^b}{\epsilon_{mn}^2}\right)\frac{1}{\epsilon_{mn}-\hbar\omega+i\eta}\nonumber\\
&\quad-\frac{e^2}{2}\int_{\bm{k}}\sum_{nm}\frac{f_{nm}s_{nm}^a(\bm{k})r_{mp}^br_{pn}^b}{\epsilon_{mn}}\left(\frac{f_{pm}}{\epsilon_{mp}-\hbar\omega-i\eta}
-\frac{f_{np}}{\epsilon_{pn}+\hbar\omega-i\eta}\right)\nonumber\\
&\quad+\frac{e^2}{2}\int_{\bm{k}}\sum_{nm}\frac{f_{nm}s_{nm}^a(-\bm{k})r_{mp}^br_{pn}^b}{\epsilon_{mn}}\left(\frac{f_{pm}}{\epsilon_{mp}-\hbar\omega+i\eta}
-\frac{f_{np}}{\epsilon_{pn}+\hbar\omega+i\eta}\right).\label{eq:shift_sym}
\end{align}
\end{widetext}
Here, we kept the terms that contribute to the leading order in $\eta$, which is $\eta^0$, and the other terms are neglected.

On the other hand, in the presence of spatial-inversion symmetry, one can take the eigenstates as $|u_{n,-\bm{k}}\rangle=\hat{P}|u_{n,\bm{k}}\rangle$, which leads to the following relations
\begin{align}
\epsilon_{nm}(-\bm{k})&=\epsilon_{nm}(\bm{k}),\label{inv1}\\
\bm{r}_{nm}(-\bm{k})&=-\bm{r}_{nm}(\bm{k}),\\
\bm{v}_{nm}(-\bm{k})&=-\bm{v}_{nm}(\bm{k}),\label{inv3}\\
O_{nm}(-\bm{k})&=(\hat{P}^\dagger\hat{O}\hat{P})_{nm}(\bm{k}).\label{inv4}
\end{align}
For instance, $O_{nm}(\bm{k})$ is an even function of $\bm k$ for spin operators.
In this case, the shift current contribution reads
\begin{widetext}
\begin{align}
\sigma_{abb}^{(2A)}(0;\omega;-\omega)
&=-\frac{i e^2}{4}\int_{\bm{k}}\sum_{nm}f_{nm}r_{mn}^b\left(-\frac{s_{nm;b}^a(\bm{k})-s_{mn;b}^a(-\bm{k})}{\epsilon_{mn}}
+\frac{\hbar (s^a_{nm}(\bm{k})+s_{mn}^a(-\bm{k}))\Delta_{mn}^b}{\epsilon_{mn}^2}\right)\nonumber\\
&\qquad\times\left[\frac{1}{\epsilon_{mn}-\hbar\omega-i\eta}+\frac{1}{\epsilon_{mn}+\hbar\omega-i\eta}\right]\nonumber\\
&\quad-\frac{e^2}{2}\int_{\bm{k}}\sum_{nm}\frac{f_{nm}[s_{nm}^a(\bm{k})+s_{nm}^a(-\bm{k})]r_{mp}^br_{pn}^b}{\epsilon_{mn}}\nonumber\\
&\qquad\times\left(\frac{f_{pm}}{\epsilon_{mp}-\hbar\omega-i\eta}
-\frac{f_{np}}{\epsilon_{pn}+\hbar\omega-i\eta}+\frac{f_{pm}}{\epsilon_{mp}+\hbar\omega-i\eta}
-\frac{f_{np}}{\epsilon_{pn}-\hbar\omega-i\eta}\right).\label{eq:shift_I}
\end{align}
\end{widetext}
From Eqs.~\eqref{inv3} and \eqref{inv4}, the spin current $s^a_{nm}$ satisfies $s^a_{nm}(\bm{k})=-s^a_{nm}(-\bm{k})$.
Therefore, the spin current in the presence of the spatial-inversion symmetry is zero.
This result for spin current reflects that spin currents are not generated in the crystal with spatial-inversion symmetry, a property that is consistent with the symmetry analysis.


\subsection{Injection current}

Next, we discuss the contribution in Eq.~\eqref{eq:inj_pre}.
Similar to $\sigma_{abc}^{(2A)}$, by rewriting the equation using velocity operator, we find that $\sigma_{abc}^{(2B)}$ is related to the injection current,
\begin{widetext}
\begin{align}
\sigma_{abc}^{(2B)}(0;\omega;-\omega)
&=\frac{e^2}{2}\int_{\bm{k}}\sum_{nmp}\left[\frac{(|O_{n\mu}||v^a_{\mu m}|e^{-i\phi^O_{n\mu}-i\phi^a_{\mu m}}+|v^a_{n\mu}||O_{\mu m}|e^{-i\phi^a_{n\mu}-i\phi^O_{\mu m}})|r_{mp}^b||r_{pn}^c|e^{-i\phi^b_{mp}-i\phi^c_{pn}}}{\epsilon_{np}}\right.\nonumber\\
&\ \ \ \ \ \left.\times\left(\frac{\rho_\beta f_{pm}}{\epsilon_{mp}-\hbar\omega_\beta-i\eta}
+\frac{\rho_\gamma f_{np}}{\epsilon_{pn}-\hbar\omega_\gamma-i\eta}\right)+(bc\beta\gamma\Leftrightarrow cb\gamma\beta)\right].\label{eq:inj}
\end{align}
\end{widetext}
Here, we expressed $\sigma_{abc}^{(2B)}$ using $\phi_{nm}$, similar to Eqs.~\eqref{phi1} and~\eqref{phi2}.
From the definitions of $\rho_\beta$ and $\rho_\gamma$, $\rho_{\beta,\gamma}\propto1/\omega_\Sigma\propto1/\eta$, which indicates that the injection current contribution is proportional to the relaxation time $1/\eta$.
To analyze the $\eta\to0$ limit, here we define 
\begin{align}
\eta_{abc}^{(2B)}&(\omega_\Sigma;\omega_\beta,\omega_\gamma)=-i\omega_\Sigma\sigma_{abc}^{(2B)}(\omega_\Sigma;\omega_\beta,\omega_\gamma),
\end{align}
which is related to the photocurrent by~\cite{Sipe2000a}
\begin{align}
\frac{d\langle J_a^{(2B)}\rangle}{dt}=\eta_{abc}^{(2B)}(\omega_\Sigma;\omega_\beta,\omega_\gamma)
E^bE^c.
\end{align}
In the presence of time-reversal symmetry, in which the relations in Eqs.~\eqref{tim1}-\eqref{tim4} holds, the injection current reads
\begin{align}
\eta_{abc}^{(2B)}&(0;\omega;-\omega)
=-\pi e^2\int_{\bm{k}}\sum_{nmp}\frac{f_{np}r_{mp}^br_{pn}^c\omega}{\epsilon_{np}}\nonumber\\
&\times
[s^a_{nm}(\bm{k})\delta(\epsilon_{np}-\hbar\omega)-s^a_{mn}(-\bm{k})\delta(\epsilon_{np}+\hbar\omega)].
\end{align}
The injection current is derived from real excitations regardless of the spin current or the orbital current.

In the presence of the spatial-inversion symmetry, on the other hand, the relations in Eqs.~\eqref{inv1}-\eqref{inv4} hold.
As a result,
\begin{align}
\eta_{abc}^{(2B)}&(0;\omega;-\omega)
=\nonumber\\
&-\frac{\pi e^2}{2}\int_{\bm{k}}\sum_{nmp}\frac{f_{np}(s^a_{nm}(\bm{k})+s_{nm}^a(-\bm{k}))\omega}{\epsilon_{np}}\nonumber\\
&\times[r_{mp}^br_{pn}^c\delta(\epsilon_{np}-\hbar\omega)-r_{mp}^cr_{pn}^b\delta(\epsilon_{np}+\hbar\omega)].
\end{align}
Similar to the shift current, the spin current is not generated if the system has the inversion symmetry.

\section{Bernevig-Hughes-Zhang model}\label{sec:BHZ}

\subsection{Model}

\begin{figure}[t]
  \includegraphics[width=\linewidth]{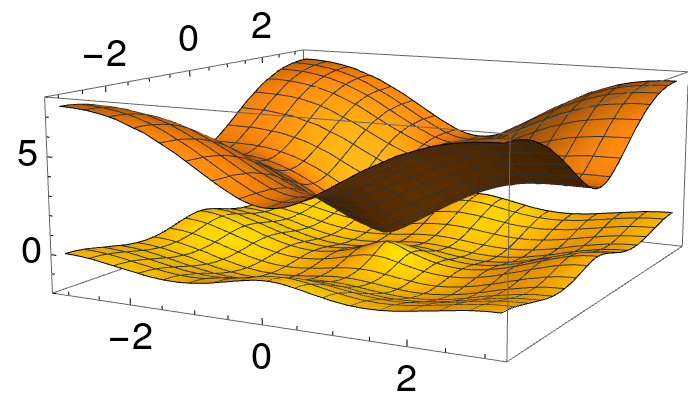}
  \caption{
  The dispersion of the BHZ model with $\epsilon_s=3.5\mathrm{eV},t_{ss}=t_{sp}=1\mathrm{eV},$ and $\epsilon_p=t_{pp}=0\mathrm{eV}$, which corresponds to $\lambda=3.5$. 
  }
  \label{fig:model}
\end{figure}

As an example, we study the nonlinear response in the Bernevig-Hughes-Zhang (BHZ) model.
The BHZ model is a popular model in the study of topological insulators, which was initially introduced as an effective model for HgTe/CdTe quantum wells~\cite{Bernevig2006a,Konig2007a}.
This model is a four band model consisting of $|s,\uparrow\rangle$, $|s,\downarrow\rangle$, $|p_x+ip_y,\uparrow\rangle$, and $|p_x-ip_y,\downarrow\rangle$ orbitals,
\begin{align}
H(\bm{k})=d_o(\bm{k})I_4+\sum_{a=1}^5d_a(\bm{k})\Gamma^a
\end{align}
where $\Gamma^a$ is the Gamma matrix which is $\Gamma^a=\{\tau_z,\tau_y,\tau_x\sigma_x,\tau_x\sigma_y,\tau_x\sigma_z\}$ and
\begin{align*}
d_0&=\frac{\epsilon_s+\epsilon_p}{2}-(t_{ss}-t_{pp})(\cos k_x+\cos k_y),\\
d_1&=\frac{\epsilon_s-\epsilon_p}{2}-(t_{ss}+t_{pp})(\cos k_x+\cos k_y),\\
d_2&=2t_{sp}\sin k_y,\\
d_3&=d_4=0,\\
d_5&=2t_{sp}\sin k_x.
\end{align*}
Here, $\tau$ and $\sigma$ are Pauli matrices acting on the orbital and spin components, respectively.

The dispersion of the BHZ model is shown in Fig.~\ref{fig:model}.
It consists of two doubly degenerate bands, which is a consequence of time- and spatial-inversion symmetries;
one of the two doubly-degenerate bands is the conduction band, and the other is the valence band.
For the parameters used in Fig.~\ref{fig:model}, which we focus on in this study, the model is known to show a topological phase transition by changing $\epsilon_s$.
It is a topological insulator at $\lambda=\epsilon_{s}/t_{ss}<4.0$ and a trivial insulator at $\lambda>4.0$; the model is a Dirac semimetal at $\lambda=4.0$.

\subsection{Orbital shift and injection currents}

\begin{figure}[tp]
  \includegraphics[width=\linewidth]{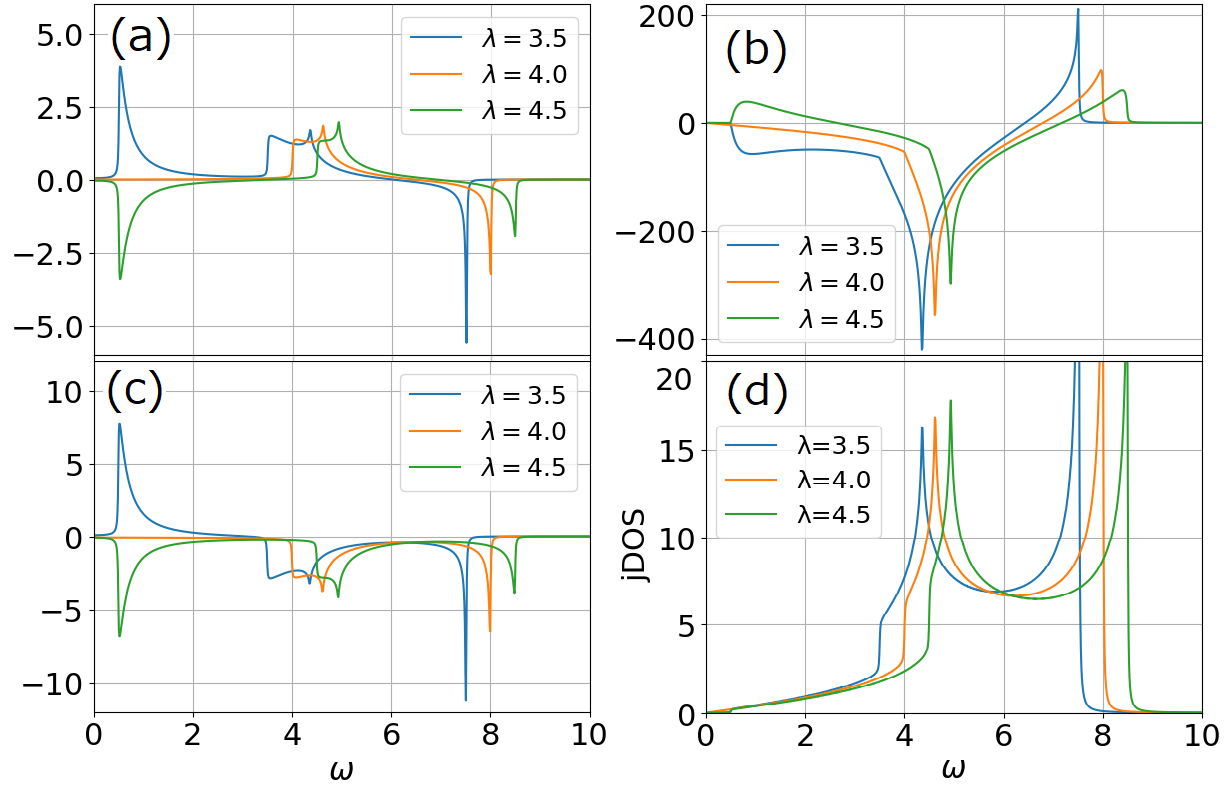}
  \caption{
  Nonlinear orbital conductivity for the BHZ model. The frequency dependence of (a) $\Re(\sigma_{xxy}^{(2A)})$, (b) $\Im(\sigma_{xxy}^{(2B)})$, and (c) $\Re(\sigma_{yyy}^{(2A)})$, and (d) joint density of states.
  }
  \label{fig:bhz}
\end{figure}

As a demonstration, here we study the orbital current for the polarization $\tau_x=\sigma_x\otimes\mathbf1$ where $\mathbf 1$ is the $2\times2$ identity matrix and $\sigma_a$ ($a=x,y,z$) are the Pauli matrices.
The same phenomenon was also studied in a recent work using a different method~\cite{Davydova2022a}, in which it was shown that $\tau_x$ is related to the electrical polarization.
Figures~\ref{fig:bhz}(a) and \ref{fig:bhz}(c) show the orbital current conductance $\sigma_{xxy}$ in Eqs.~\eqref{eq:shift} and \eqref{eq:inj}.
Unlike the photocurrent, this orbital response occurs in centrosymmetric materials.
As pointed out in a previous work~\cite{Davydova2022a}, the sign of the response coefficient is the opposite in the trivial ($\lambda=3.5$) and topological ($\lambda=4.5$) cases.
However, unlike the previous study, the results in Figs.~\ref{fig:bhz}(a) and~\ref{fig:bhz}(c) show a peak-like structure at the lower-frequency edge.
This is merely a consequence of the difference in the definition of $\sigma_{xxy}^{(2)}$;
we defined $J_x=\sigma_{xxy}^{(2)}E_xE_y$ whereas $J_x=\sigma_{xxy}^{(2)}A_xA_y$ was used in the previous work.
Here, $\bm{A}$ is a vector potential.
In addition, the wave numbers $k_x$ and $k_y$ in this paper correspond to $k_y$ and $k_x$ in the previous work, which means $\sigma_{yyy}$ in Fig.~\ref{fig:bhz} corresponds to $\sigma_{xxx}$ in the previous work.
Except for the difference in the definition, the nonlinear response theory reproduces the results for linearly-polarized light obtained using the Floquet-Keldysh formalism.

In fact, for the case of four-band models, Eq.~\eqref{eq:shift_I} is equivalent to the one obtained by the Keldysh-Floquet method.
To show this, we focus on a four-band system with two valence and two conduction bands.
For concreteness, we label the valence bands $1$ and $2$ and the conduction bands $3$ and $4$.
For the shift current contribution by linearly polarized light ($b=c$), the term containing $s^a_{nm;c}$ is dominant and $s^a_{nm}(-\bm{k})=-s^a_{mn}(\bm{k})$ holds in the presence of the time-reversal symmetry.
With these conditions, $\sigma_{abb}^{(2A)}$ reads
\begin{align}
\sigma_{abb}^{(2A)}&=-\frac{e^2}{2\hbar^2\omega^2}\int\frac{d^2k}{(2\pi)^2}\sum_{\substack{n=1,2\\m=3,4}}\nonumber\\
&\qquad(v_{mn}^bs_{nm;b}^a-s_{mn;b}^av^b_{nm})\delta(\epsilon_{31}-\hbar\omega).\label{eq:BHZ-sigma}
\end{align}
Note that $\epsilon_1(\bm k)=\epsilon_2(\bm k)$ and $\epsilon_3(\bm k)=\epsilon_4(\bm k)$ in the current case.
Equation~\ref{eq:BHZ-sigma} is equivalent to Eq.~(4) in Ref.~\cite{Davydova2022a}. 
Therefore, our result in Eq.~\eqref{eq:shift} is a generalization to generic non-interacting models.

In addition to the shift current contribution, we find that the injection current also occurs in the BHZ model.
Figure~\ref{fig:bhz}(b) shows the result of $\Im(\sigma^{(2B)}_{xxy})$.
The result shows that the sign of $\Im(\sigma^{(2B)}_{xxy})$ in the low-frequency region also changes by the topological phase transition, similar to the case of the shift current contribution.

The origin of sign reversal is, however, different.
For the case of the injection current, it is a consequence of the fact that the sign of the imaginary part of $r^b_{mp}r^c_{pn}$ in Eq.~\eqref{eq:inj_pre} changes in the $|\bm{k}|=k\to0$ limit.
To understand its origin, we consider the BHZ Hamiltonian near the $\Gamma$ point, which reads $H_0(\bm k)=k_x\sigma_x+k_y\sigma_y+m\sigma_z$, where $m=\epsilon_s-4$ and $H(\bm{k})=\mathrm{diag}(H_0(\bm{k}),H_0(-\bm{k})^*)$ in the basis $(|s,\uparrow\rangle,|p_x+ip_y,\uparrow\rangle,|s,\downarrow\rangle,|p_x-ip_y,\downarrow\rangle)$.
For the case with two valence and two conduction bands, as in Fig.~\ref{fig:model}, the eigenstates of $H_0$ are
\begin{align}
&|1,\bm{k}\rangle=\frac{1}{2mk}\left(\begin{array}{c}
ik\\2m(k_x-ik_y)
\end{array}\right),\nonumber\\
&|2,\bm{k}\rangle=\frac{1}{2m}\left(\begin{array}{c}
-2im\\k_x-ik_y
\end{array}\right),\label{eq:dirac1}
\end{align}
for $m>0$ and
\begin{align}
&|1,\bm{k}\rangle=\frac{1}{2m}\left(\begin{array}{c}
-2im\\k_x-ik_y
\end{array}\right),\nonumber\\
&|2,\bm{k}\rangle=-\frac{1}{2mk}\left(\begin{array}{c}
ik\\2m(k_x-ik_y)
\end{array}\right),\label{eq:dirac2}
\end{align}
for $m<0$, where $k^2=k_x^2+k_y^2$.
By using these eigenstates, we find that $r^x_{12}r^y_{21}=\text{sgn}(m)/(4m^2)$ near the $\Gamma$ point.
Therefore, the sign of $\Im(\sigma^{(2B)}_{xxy})$ depends on $m$.
However, we note that the injection current corresponds to the imaginary part, and therefore, is different from the shift current.

\begin{figure}[tp]
  \includegraphics[width=\linewidth]{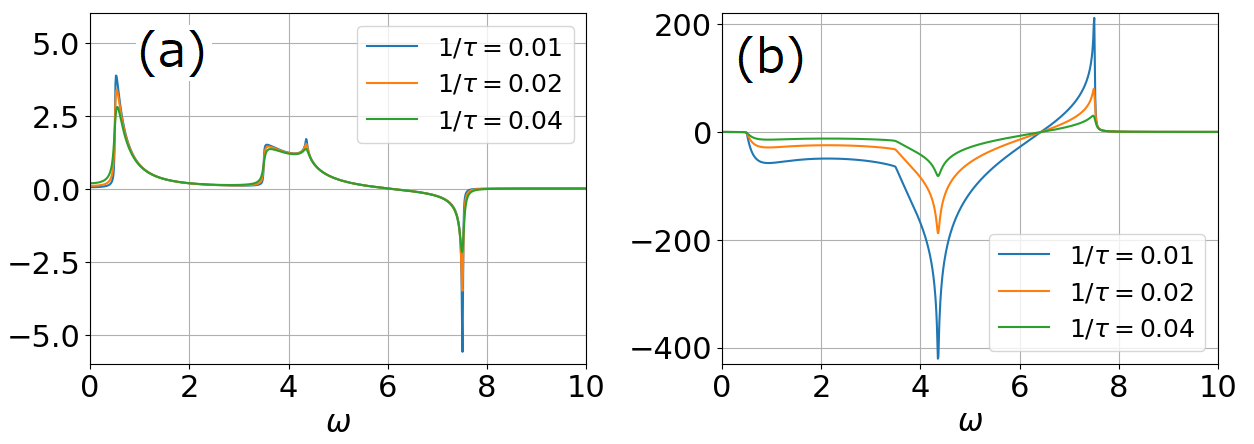}
  \caption{The $\omega$ dependence of (a) $\Re(\sigma_{xxy}^{(2A)})$ and (b) $\Im(\sigma_{xxy}^{(2B)})$ with different relaxation time $\tau$.}
  \label{fig:bhz_delta}
\end{figure}

Next, we look into the relaxation time dependence. 
As shown in Fig.~\ref{fig:bhz_delta}(a), the real part of $\sigma_{xxy}^{(2A)}$ does not depend on the relaxation time, similar to the electrical shift current. 
On the other hand, the imaginary part of $\sigma_{xxy}^{(2B)}$ in Fig.~\ref{fig:bhz_delta}(b) increases linearly with the relaxation time, similar to the injection current.
The results show that the relaxation-time dependence is the same as that of the electrical photocurrents.

\subsection{Four-band effect}

The discussion up to here is essentially that of two-band systems, as implied from Eqs.~\eqref{eq:dirac1} and \eqref{eq:dirac2}.
To study effects beyond two-band models, we consider a Rashba term
\begin{align}
    H'(\bm{k})=E(\sin(k_x)\sigma_y-\sin(k_y)\sigma_x),
\end{align}
which occurs as a result of inversion symmetry being broken by the substrate or a gate bias.

\begin{figure}[tp]
  \includegraphics[width=\linewidth]{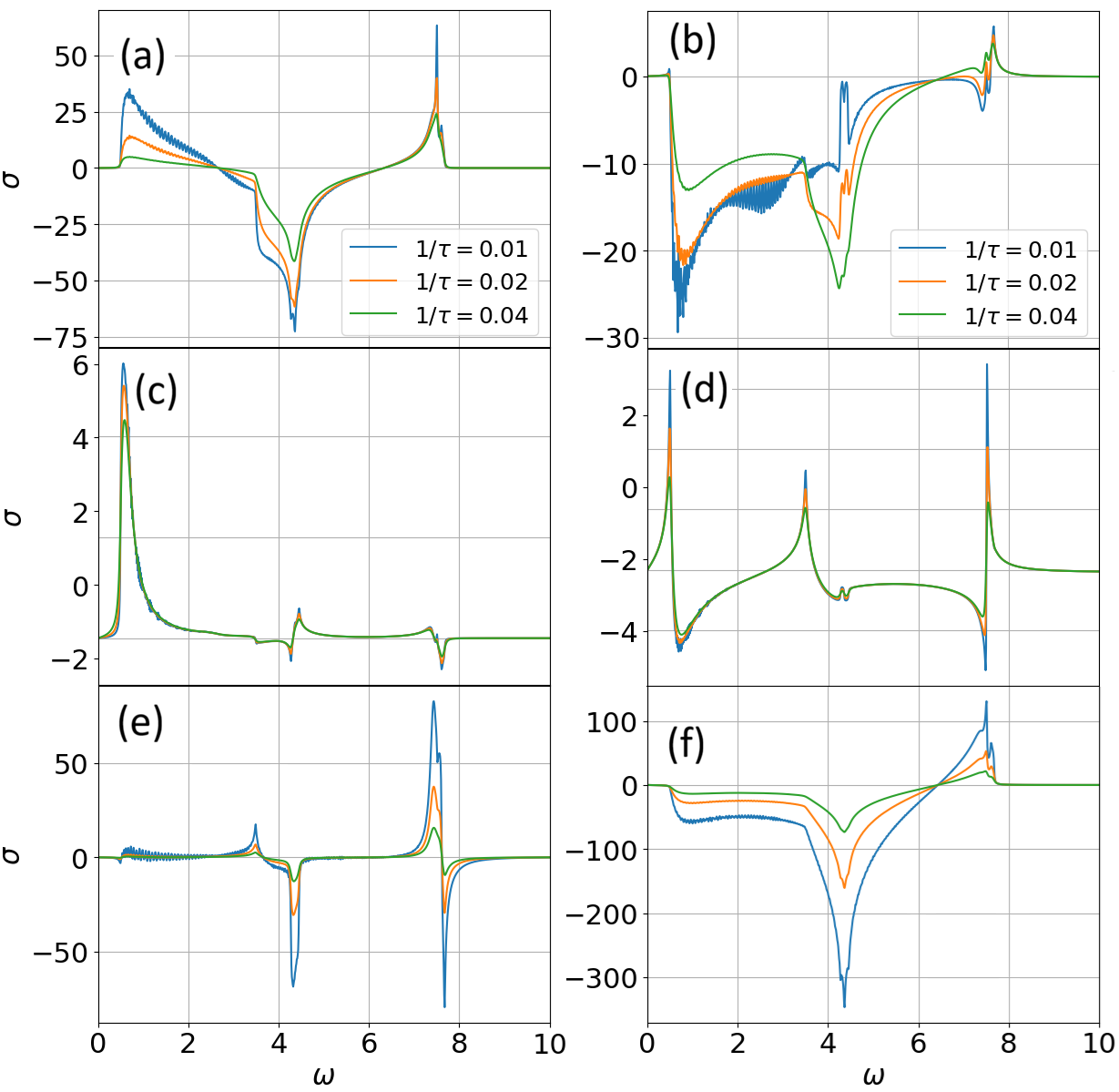}
  \caption{
  The imaginary part of nonlinear (a),(c),(e) spin and (b),(d),(f) orbital conductivities for the BHZ model with the Rashba term.
  (a),(b) Nonlinear conductivities calculated with Eq.~\eqref{2nd}, (c),(d) their shift-current part in Eq.~\eqref{eq:shift}
  and (e),(f) their injection-current part in Eq.~\eqref{eq:inj}.
  The results are for $\epsilon_s=3.5\mathrm{eV},t_{ss}=t_{sp}=1\mathrm{eV}$, $\epsilon_p=t_{pp}=0\mathrm{eV}$, and $E=0.1\mathrm{eV}$.
  }
  \label{fig:compbhz1}
\end{figure}

With $H'$, the symmetry allows spin photocurrent, $\hat O=\sigma_a$ ($a=x,y,z$), in addition to the orbital current.
Figures~\ref{fig:compbhz1}(a) and~\ref{fig:compbhz1}(b) show the imaginary part of the $\sigma_x$ spin and orbital conductivities $\sigma^{(2)}_{xxy}$. 
Unlike the results shown in Fig.~\ref{fig:bhz_delta}, those shown in Fig.~\ref{fig:compbhz1} are not proportional to the relaxation time; this behavior can be attributed to the existence of both shift and injection current contributions.
Using the injection current formula in Eq.~\eqref{eq:inj}, we can obtain the shift and injection current contributions, as shown in Figs.~\ref{fig:compbhz1}(c) and~\ref{fig:compbhz1}(d).
This is contrasting to the cases without the Rashba term, in which the shift and injection current contributions correspond to the real and imaginary parts, respectively.
The manifestation of both shift and injection current contributions is presumably related to the lack of inversion symmetry.

\begin{figure}[tp]
  \includegraphics[width=\linewidth]{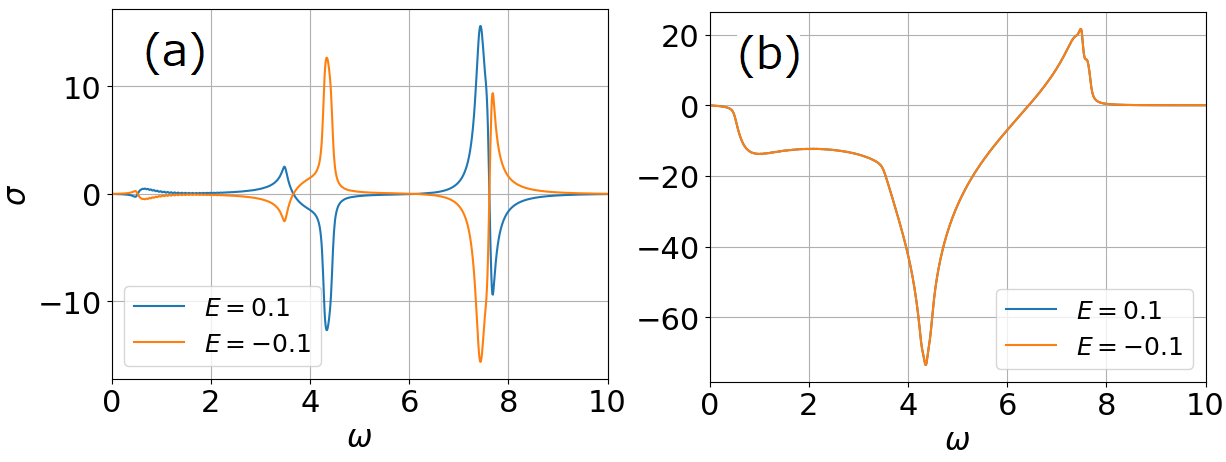}
  \caption{
  (a) spin and (b) orbital conductivities $\mathrm{Im}\sigma^{(2B)}_{xxy}$, where $\tau=0.4$.
  }
  \label{fig:compbhz3}
\end{figure}

Next, we calculate the $E$ dependence of the conductivities. 
Figure~\ref{fig:compbhz3} shows the results for spin and orbital current, respectively.
They suggest that the spin current reverses direction depending on the sign of the Rashba term, whereas the orbital current does not.
This behavior is anticipated in accordance with the symmetry argument.
The sign reversal of the Rashba term is analogous to the inversion operation in Eq.~\eqref{tim4}; algebraically, the following 
relation holds, $O_{nm}(\bm{k},E)=(\hat{\sigma}_y\hat{O}^*\hat{\sigma}_y)_{mn}(-\bm{k},-E)$.
Thereby, only the matrix elements of the spin current operator change sign, and so does the spin current conductivity.

To understand the sign reversal without gap closing, which was not the case for the orbital current, we next look into the formal structure of the conductivity formula.
To this end, we focus on the velocity operator arising from the Rashba term as $\bm{v}'=\frac1\hbar\partial_{\bm k}H'(\bm{k})$, which reads $(v'_x(\bm{k}),v'_y(\bm{k}))=(E\cos(k_x)\sigma_y,-E\cos(k_y)\sigma_x)$.
For example, the spin current operator for the $\sigma_x$ spin current is
\begin{align}
\frac{1}{2}\{\sigma_x,v_x'\}_{nm}&=0,\\
\frac{1}{2}\{\sigma_x,v_y'\}_{nm}&=-E\cos(k_y)\delta_{nm},
\end{align}
in the eigenstate basis.
It shows that the Rashba term induces a velocity component, which is related to the Rashba splitting, and contributes to the photocurrents related to the diagonal component of the velocity operator.

As a result, new properties such as non-zero spin current and a sign change without a band inversion occurs by adding the Rashba term to the BHZ Hamiltonian.
This phenomenon can be rephrased as follows.
When considering the Hamiltonian without the Rashba term, the matrix can be block-diagonalized into the $2\times2$ matrix,
with the two eigenstates corresponding to the up and the down spin.
In this case, $\sigma^x_{nm}(\bm{k})$ is not zero when $n$ and $m$ indicate the opposite spins, while $\bm{r}_{nm}(\bm{k})$ or $\bm{v}_{nm}(\bm{k})$ is not zero when $n$ and $m$ are the same spin states.
Therefore, the conductivity for $\hat{O}=\hat{\sigma}_x$ is zero.
In contrast, when we introduce the Rashba term, the Hamiltonian can no longer be block-diagonalized, and the eigensates are superposition of the up and down spin components.
In such a case, $\sigma^x_{nm}(\bm{k})$ generally becomes nonzero for all $n$ and $m$, allowing a finite spin current.

\section{Luttinger model}\label{sec:lutinger}

\subsection{Model}

As an example of orbital conductivity related to orbital magnetization~\cite{bernevig2005a,Wang2024}, we study the optical conductivity and orbital current conductivity of the two-dimensional Luttinger model~\cite{Luttinger1956} with uniaxial strain.
This model is an effective model for the films of Pr$_2$Ir$_2$O$_7$~\cite{Moon2013a,Terasawa2024a} and $\alpha$-Sn~\cite{Zhang2018,Anh2021a}. 
This model is known to become a topological insulator in the presence of a uniaxial anisotropy~\cite{Moon2013a,Zhang2018,Anh2021a,Duan2022a}.
Therefore, it is another model that shows a topological phase transition.

The Luttinger model is a four-band model consisting of orbitals $|J,m_j\rangle=|\frac{3}{2}$, $\frac{3}{2}\rangle,|\frac{3}{2},\frac{1}{2}\rangle$, $|\frac{3}{2},-\frac{1}{2}\rangle$, and $|\frac{3}{2},-\frac{3}{2}\rangle$.
The Hamiltonian is given by
\begin{align}
H(\bm{k})=&\frac{\gamma_1}{2}k^2+\gamma_2\left(\frac{5}{4}k^2-\sum_{\alpha=x,y} k_a^2J_a^2\right)\nonumber\\
&\ \ \ \ \ 
-\gamma_3k_xk_y\{J_x,J_y\}-\Delta\left(J_z^2-\frac{5}{4}\right),\label{eq:lut}
\end{align}
where $\gamma_{1,2,3}$ are the Luttinger parameters, $J_{x,y,z}$ are the $S=3/2$ spin operators,
\begin{align}
J_x&=\left(\begin{array}{cccc}
0&\frac{\sqrt{3}}{2}&0&0\\
\frac{\sqrt{3}}{2}&0&1&0\\
0&1&0&\frac{\sqrt{3}}{2}\\
0&0&\frac{\sqrt{3}}{2}&0
\end{array}\right),\\
J_y&=\left(\begin{array}{cccc}
0&-\frac{\sqrt{3}}{2}i&0&0\\
\frac{\sqrt{3}}{2}i&0&-i&0\\
0&i&0&-\frac{\sqrt{3}}{2}i\\
0&0&\frac{\sqrt{3}}{2}i&0
\end{array}\right),\\
J_z&=\left(\begin{array}{cccc}
\frac{3}{2}&0&0&0\\
0&\frac{1}{2}&0&0\\
0&0&-\frac{1}{2}&0\\
0&0&0&-\frac{3}{2}
\end{array}\right),
\end{align}
and $\Delta$ is the uniaxial strain. 

\begin{figure}
  \includegraphics[width=\linewidth]{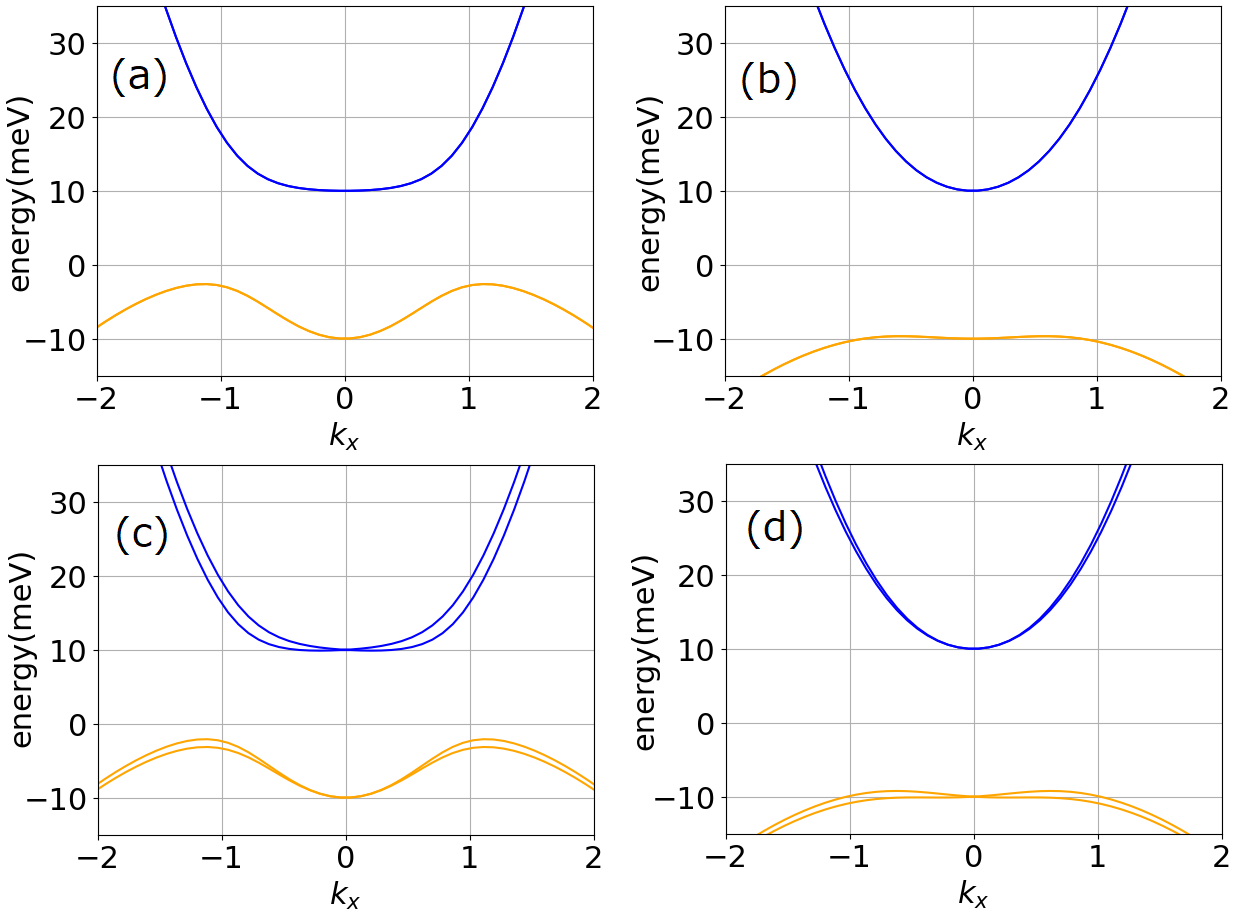}
  \caption{
  The dispersion of the Luttinger model along the $k_y=0$ line for (a) $\Delta=10$ meV and $E=0$ meV, (b) $\Delta=-10$ meV and $E=0$ meV, (c) $\Delta=10$ meV and $E=1$ meV, and (d) $\Delta=-10$ meV and $E=1$ meV. 
  The other parameters are $\gamma_1=14.97$ meV, $\gamma_2=10.61$ meV, and $\gamma_3=8.52$ meV.
  }
  \label{fig:model_lut}
\end{figure}

The dispersion of the Luttinger model is shown in Fig.~\ref{fig:model_lut}.
This model consists of two doubly degenerate bands, reflecting the 
time- and spatial-inversion symmetries. 
In this study, we focus on the cases with two valence and two conduction bands.
The valence and conduction bands touch at $\bm k=\bm0$ in the absence of the uniaxial strain ($\Delta=0$), and a gap exists when $\Delta\ne0$;
it is a topological insulator for $\Delta>0$ and a trivial insulator for $\Delta<0$.
The realization of such a topological phase transition was recently reported in $\alpha$-Sn by tuning the thickness of thin films~\cite{Anh2021a}.

In addition, we consider a Rashba term 
\begin{align}
H'(\bm{k})=E(k_xJ_y-k_yJ_x),
\end{align}
similar to the BHZ model.
This term breaks the spatial-inversion symmetry, and as a consequence, it removes the spin degeneracy of the degenerate bands as shown in Fig.~\ref{fig:model_lut}(c) and \ref{fig:model_lut}(d).
In addition, the inversion symmetry breaking allows for a nonzero orbital current associated with the orbital angular momentum.
Therefore, the Luttinger model with Rashba term is an interesting model for studying the nonlinar orbital current.

\subsection{Optical conductivity}

\begin{figure}
  \includegraphics[width=\linewidth]{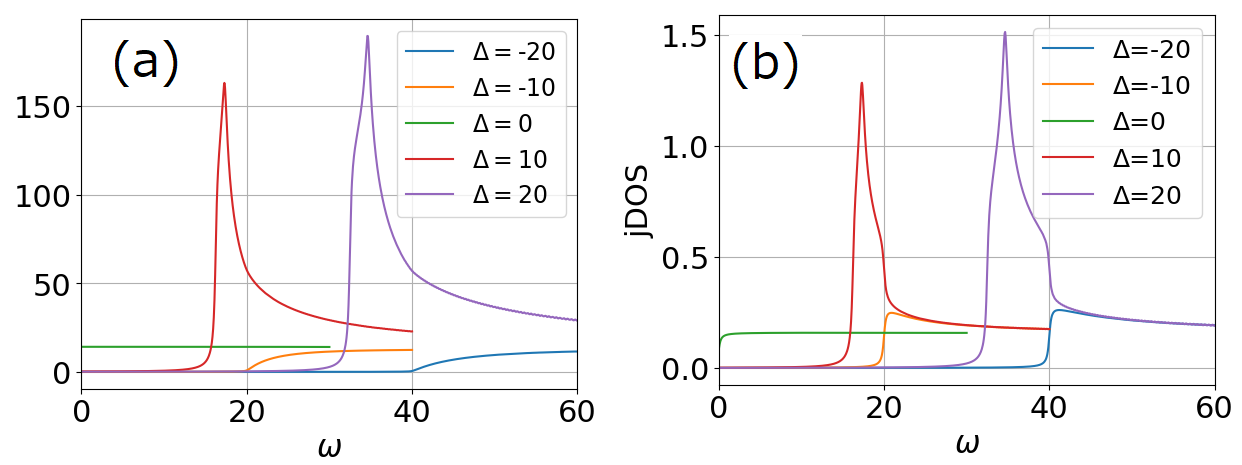}
  \caption{(a) The optical conductivity $\Re(\sigma_{xx})$ and (b) joint density of states of the Luttinger model.
  The cutoff $\Lambda$ is 30, 40, and 60 for $|\Delta|=$0, 10, and 20, respectively.
  }
  \label{fig:linearlut}
\end{figure}

To study how the topological transition manifests in the optical properties, we first look into the optical conductivity.
Figure~\ref{fig:linearlut} shows the optical conductivity of the Luttinger model for different $\Delta$.
For the calculation of optical conductivity, we set the cutoff $\Lambda$ by the difference in energy between the conduction and valence bands at the same $\bm k$. 
As seen in Fig.~\ref{fig:linearlut}, a huge difference appears between the topological and trivial phases.
In the topological phases ($\Delta>0$), the optical conductivity resembles the jDOS in Fig.~\ref{fig:linearlut}(b).
On the other hand, the optical conductivity for trivial phases ($\Delta<0$) shows a smooth monotonic increase with respect to $\omega$, which is different from that of jDOS.

To understand the origin of the unique behavior at $\Delta<0$, we derive the analytic form of the $\omega$ dependence of optical conductivity around $\omega=2|\Delta|$.
In this case, only the electrons near the $\Gamma$ point contribute to the optical conductivity.
By expanding the dispersion about the $\Gamma$ point up to the second order in $\bm k$,
\begin{align}
H_{kp}&=\mathrm{diag}(-\Delta +\frac{\gamma_1+\gamma_2}2k^2 
,
\Delta+\frac{\gamma_1-\gamma_2}2k^2
,\nonumber\\
&\ \ \ \ \ \ \ \ \Delta+\frac{\gamma_1-\gamma_2}2k^2
,
-\Delta +\frac{\gamma_1+\gamma_2}2k^2).
\end{align}
It corresponds to neglecting the matrix elements between the valence-band and conduction-band states at $\bm k=\bm0$.
With this approximation, we can calculate the optical conductivity of the linear response by substituting the eigenvalues and eigenstates of $H_{kp}$ into Eq. \eqref{eq:1st},
\begin{align}
\sigma_{xx}^{(1)}(\omega)
&=\frac{3}{4}\left(1+\frac{\gamma_3^2}{\gamma_2^2}\right)\frac{\omega+2\Delta}{\omega}.
\end{align}
Here, we kept the off-diagonal terms of the velocity matrix as it gives the leading order contribution to $\sigma^{(1)}_{xx}$, i.e., we used $H$ instead of $H_{kp}$ to calculate $v^x_{nm}$.
This matches the behavior of the optical response of the low energy limit ($\omega=2\Delta$).
On the other hand, Fig.~\ref{fig:model_lut}(a) shows that the band bottom of the conduction band is located away from the $\Gamma$ point.
The wine-bottle structure of the electronic bands in the trivial phase enhances the jDOS as shown in Fig.~\ref{fig:linearlut}(b), which is reflected in the optical conductivity.
Measurement of different $\omega$ dependencies in optical conductivity allows distinguishing trivial and topological phases.

\subsection{Nonlinear orbital current}

Next, we look at the second-order orbital current related to the $J=3/2$ moment of the Luttinger model.
Unlike the case of the BHZ model, the orbital current here is related to the angular momentum.
In this sense, the orbital current resembles that of the spin current~\cite{Maekawa2012a}.
However, in the Luttinger model, the spin and orbital angular momenta are strongly entangled due to the spin-orbit interaction.
Hence, it is an orbital current.
Additionally, unlike the orbital photocurrent in the BHZ model — which essentially describes the physics of a two-band model — effects beyond two-band physics may emerge as the pair of conduction and valence bands in the Luttinger model are entangled.

\begin{figure}[tbp]
  \includegraphics[width=0.85\linewidth]{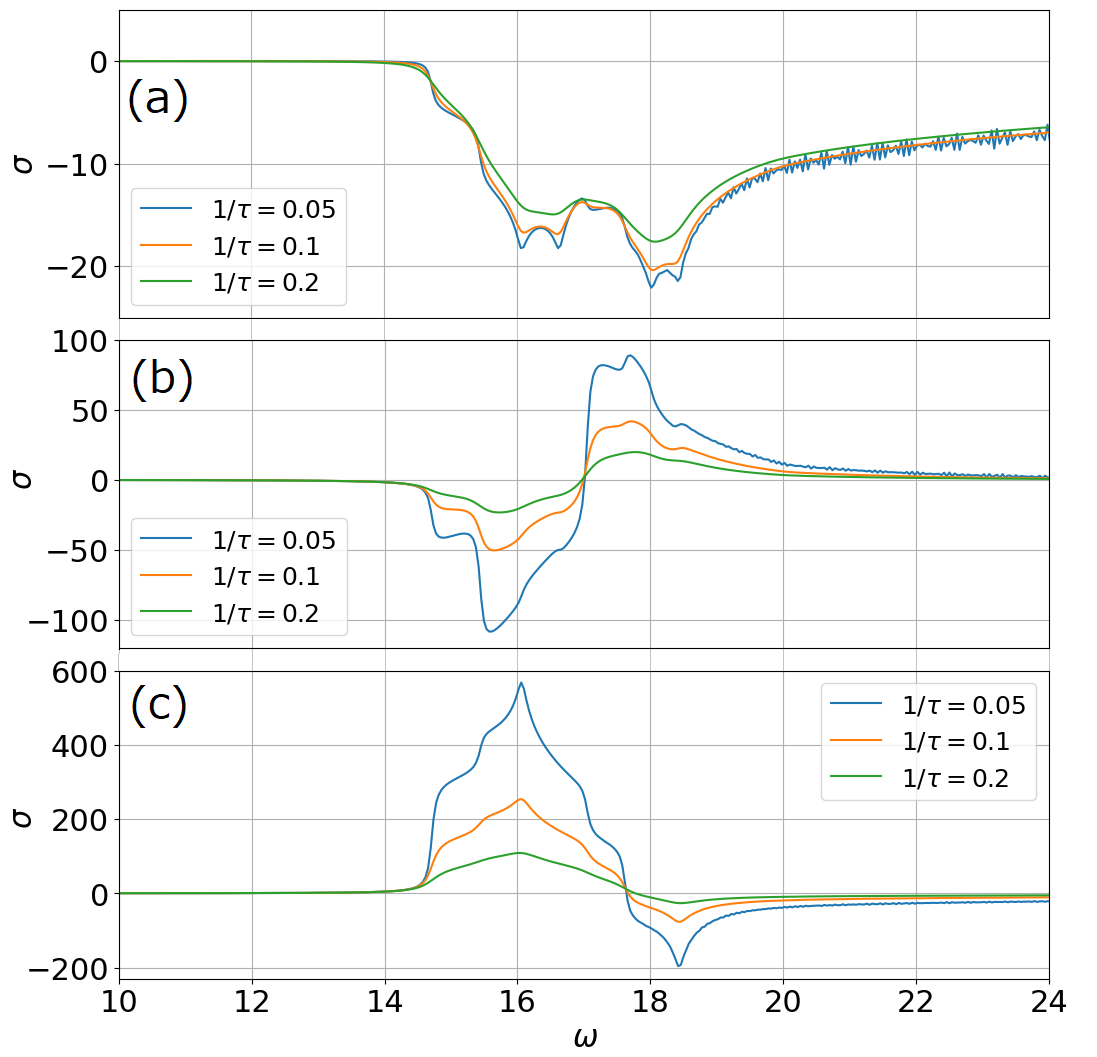}
  \caption{
  Nonlinear orbital conductivity of the Luttinger model, (a) $\Im(\sigma_{xxy}^{shift}(\omega))$, (b) $\Re(\sigma_{xxy}^{inj}(\omega))$, and (c) $\Re(\sigma_{yyy}^{inj}(\omega))$.
  The results are for $\Delta=10$ meV and $E=1$ meV.
  }
  \label{fig:2ndlut}
\end{figure}

The results of the orbital current in the Luttinger model are shown in Fig.~\ref{fig:2ndlut}.
The orbital current does not occur for the $E=0$ case, as the inversion symmetry prohibits the orbital photocurrent.
Therefore, we focus on the cases with $E=1$, which is shown in Fig. \ref{fig:2ndlut}.

Regarding the relaxation-time dependence, unlike the case of photocurrent, we find that the real part of $\sigma_{xxy}$ and $\sigma_{yyy}$ increases linearly with $\tau$ whereas the imaginary part converges to a finite value at $\tau\to\infty$.
It shows that the orbital current by a linearly-polarized light behaves like an injection current in terms of the relaxation time dependence, and the polarization-dependent part of the response to a circularly-polarized light is inert to the relaxation time of electrons.

\begin{figure}[tbp]
  \includegraphics[width=\linewidth]{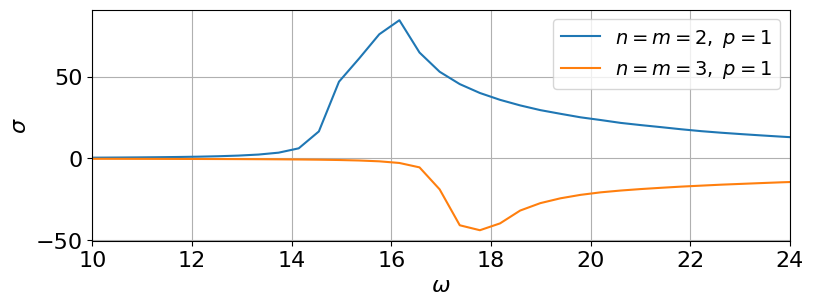}
  \caption{Orbital conductivity $\Re(\sigma_{yyy}^{inj})$ 
  where $E=1$meV and $\tau=0.2$.}
  \label{fig:lut_index}
\end{figure}
\begin{figure}[htbp]
  \includegraphics[width=\linewidth]{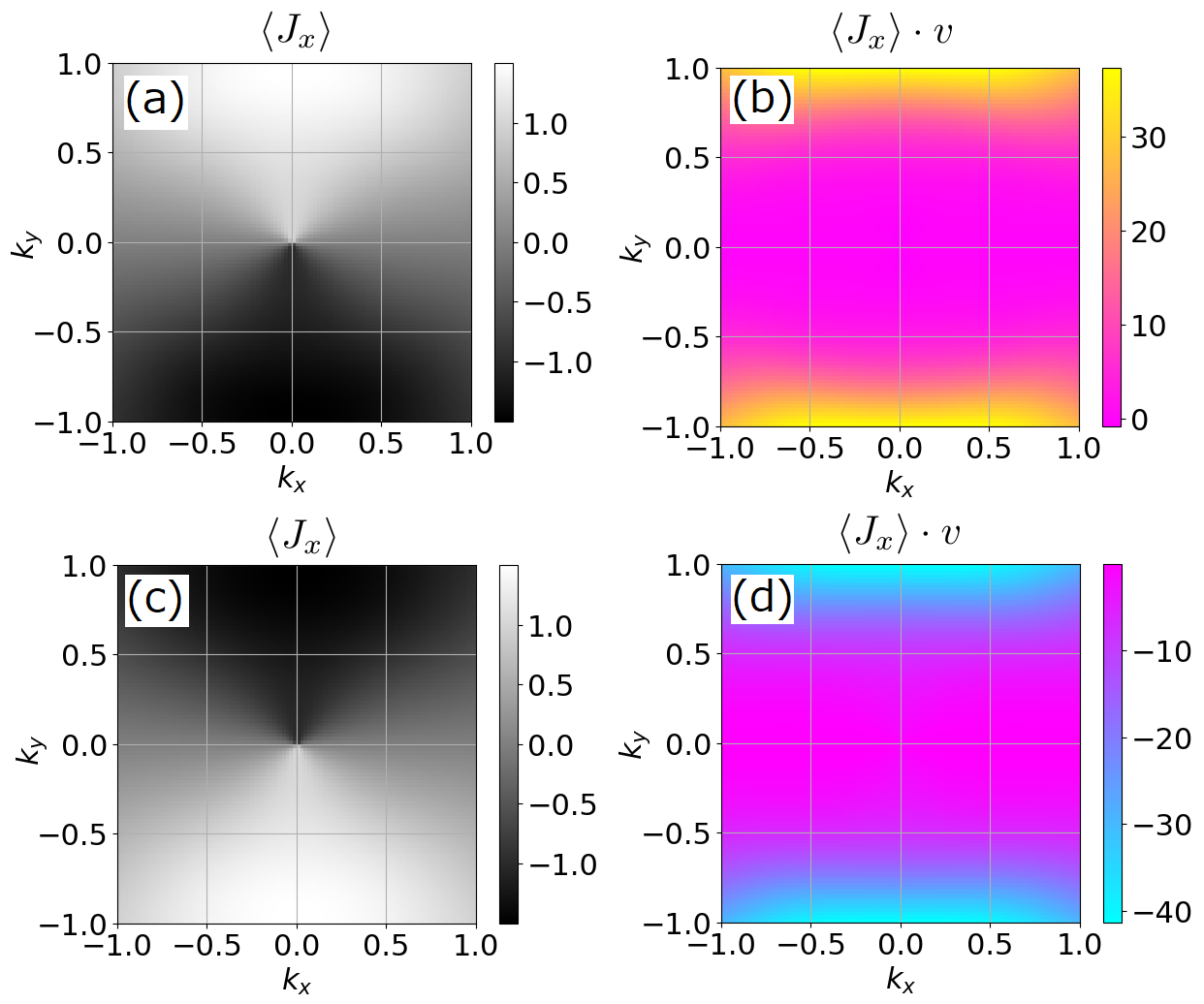}
  \caption{
  Values of $\langle J_x\rangle$ (a),(c) and $\langle J_x\rangle\cdot \bm{v}$ (b),(d) in $\bm{k}$-space.
  (a), (b) and (c),(d) mean band-2 and band-3 respectively.}
  \label{spin_kspace}
\end{figure}

For the real part of $\sigma_{xxy}$ and $\sigma_{yyy}$, we also find a sign change of the conductivity near  $\omega=17$ meV, as in Figs.~\ref{fig:2ndlut}(b) and \ref{fig:2ndlut}(c).
It can be easily understood by considering the positive and negative terms respectively.
In fact, these graphs represent the sum of positive and negative peaks, each located at a different $\omega$ (Fig. \ref{fig:lut_index}.
Here, we named band-0, 1 the valence bands, and band-2, 3 the conduction bands.
These two peaks follow the joint density of states.

Finally, we consider the reason why these two graphs in Fig. \ref{fig:lut_index} exhibit opposite signs.
It is explained by considering the orbital current $\bm{s}_{nn}$ as the product of the expectation value of the spin operator and the group velocity $\langle J_x\rangle\cdot \bm{v}$ of the band.
Figure~\ref{spin_kspace} shows that the expectation values of the spin operator for bands 2 and 3 have opposite signs,
 and consequently the expectation values of the orbital current, $s^y_{22}$ and $s^y_{33}$, also exhibit opposite signs.
Here, the injection current formula is proportional to $s^a_{nn}$ as shown in Eq.~\eqref{inv1}, so focusing on the contributions from bands 2 and 3, the orbital conductivities $\Re(\sigma_{yyy}^{(2B)})$ corresponding to these bands have opposite signs.
This explains why one of the graphs in Fig. \ref{fig:lut_index} has a positive value, while the other has a negative value.


\section{Summary and Discussion}\label{sec:summary}

In this paper, we studied the properties of linear optical conductivity and the second-order response of orbital currents.
We derived a general formula to calculate the second-order conductivity for orbital current which is a generalization of the formalism developed by Sipe \textit{et al.}~\cite{Sipe2000a}. 
The formula, expressed in terms of Lehmann representation, is straightforwardly applicable to complex models including those for real materials.
By applying the theory to Bernivig-Hughes-Zhang and Luttinger models, we find that there are shift and injection current contributions to the orbital conductivity for electric polarization and orbital magnetization.
In addition, we find that the relaxation time dependence of the orbital conductivity is somewhat distinct from that of the photocurrent.
For the orbital current for polarization, the response to the linearly-polarized light is inert to the relaxation time while that for the circularly-polarized light increases linearly with the relaxation time.
On the other hand, for the spin current for orbital magnetization, the relaxation time dependence is the opposite for the linearly- and circularly-polarized lights.
Moreover, we find that the frequency dependence of the optical conductivity of the linear response differs significantly between the trivial and topological phases in Luttinger model.
This suggests that measuring the frequency dependence of the  conductivity may provide a method to distinguish between the trivial and topological phases.

Recently, spin and orbital currents have attracted attention for application purposes~\cite{Dong2025,Wang2024}, in which the orbital degree of freedom is expected to be a new carrier of information.
The formulas presented in this paper can be used to calculate the second-order response of spin and orbital currents in various materials.
In particular, the wide applicability of the formalism used in this work enables quantitative calculation of the spin and orbital current in complex models for real materials, thereby, contributes to finding materials suitable for measuring spin and orbital currents.

\acknowledgements
We thank R. Matsunaga for fruitful discussions.
This work is supported by JSPS KAKENHI (Grant No. JP18H04222, JP19K14649, JP23K03275, and JP25H00841) and JST PRESTO (Grant No. JPMJPR2452).


\begin{thebibliography}{44}%
\makeatletter
\providecommand \@ifxundefined [1]{%
 \@ifx{#1\undefined}
}%
\providecommand \@ifnum [1]{%
 \ifnum #1\expandafter \@firstoftwo
 \else \expandafter \@secondoftwo
 \fi
}%
\providecommand \@ifx [1]{%
 \ifx #1\expandafter \@firstoftwo
 \else \expandafter \@secondoftwo
 \fi
}%
\providecommand \natexlab [1]{#1}%
\providecommand \enquote  [1]{``#1''}%
\providecommand \bibnamefont  [1]{#1}%
\providecommand \bibfnamefont [1]{#1}%
\providecommand \citenamefont [1]{#1}%
\providecommand \href@noop [0]{\@secondoftwo}%
\providecommand \href [0]{\begingroup \@sanitize@url \@href}%
\providecommand \@href[1]{\@@startlink{#1}\@@href}%
\providecommand \@@href[1]{\endgroup#1\@@endlink}%
\providecommand \@sanitize@url [0]{\catcode `\\12\catcode `\$12\catcode
  `\&12\catcode `\#12\catcode `\^12\catcode `\_12\catcode `\%12\relax}%
\providecommand \@@startlink[1]{}%
\providecommand \@@endlink[0]{}%
\providecommand \url  [0]{\begingroup\@sanitize@url \@url }%
\providecommand \@url [1]{\endgroup\@href {#1}{\urlprefix }}%
\providecommand \urlprefix  [0]{URL }%
\providecommand \Eprint [0]{\href }%
\providecommand \doibase [0]{http://dx.doi.org/}%
\providecommand \selectlanguage [0]{\@gobble}%
\providecommand \bibinfo  [0]{\@secondoftwo}%
\providecommand \bibfield  [0]{\@secondoftwo}%
\providecommand \translation [1]{[#1]}%
\providecommand \BibitemOpen [0]{}%
\providecommand \bibitemStop [0]{}%
\providecommand \bibitemNoStop [0]{.\EOS\space}%
\providecommand \EOS [0]{\spacefactor3000\relax}%
\providecommand \BibitemShut  [1]{\csname bibitem#1\endcsname}%
\let\auto@bib@innerbib\@empty
\bibitem [{\citenamefont {Nelson}(2003)}]{Nelson2003a}%
  \BibitemOpen
  \bibfield  {author} {\bibinfo {author} {\bibfnamefont {Jenny}\ \bibnamefont
  {Nelson}},\ }\href@noop {} {\emph {\bibinfo {title} {The Physics of Solar
  Cells}}},\ Properties of Semiconductor Materials\ (\bibinfo  {publisher}
  {Imperial College Press},\ \bibinfo {address} {London},\ \bibinfo {year}
  {2003})\ p.\ \bibinfo {pages} {384}\BibitemShut {NoStop}%
\bibitem [{\citenamefont {Würfel}(2005)}]{Wurfel2005a}%
  \BibitemOpen
  \bibfield  {author} {\bibinfo {author} {\bibfnamefont {Peter}\ \bibnamefont
  {Würfel}},\ }\href {\doibase 10.1002/9783527618545} {\emph {\bibinfo {title}
  {Physics of Solar Cells: From Principles to New Concepts}}}\ (\bibinfo
  {publisher} {Wiley-VCH},\ \bibinfo {address} {Weinheim},\ \bibinfo {year}
  {2005})\BibitemShut {NoStop}%
\bibitem [{\citenamefont {von Baltz}\ and\ \citenamefont
  {Kraut}(1981)}]{vBaltz1981a}%
  \BibitemOpen
  \bibfield  {author} {\bibinfo {author} {\bibfnamefont {Ralph}\ \bibnamefont
  {von Baltz}}\ and\ \bibinfo {author} {\bibfnamefont {Wolfgang}\ \bibnamefont
  {Kraut}},\ }\bibfield  {title} {\enquote {\bibinfo {title} {Theory of the
  bulk photovoltaic effect in pure crystals},}\ }\href {\doibase
  10.1103/PhysRevB.23.5590} {\bibfield  {journal} {\bibinfo  {journal} {Phys.
  Rev. B}\ }\textbf {\bibinfo {volume} {23}},\ \bibinfo {pages} {5590--5596}
  (\bibinfo {year} {1981})}\BibitemShut {NoStop}%
\bibitem [{\citenamefont {Belinicher}\ \emph {et~al.}(1982)\citenamefont
  {Belinicher}, \citenamefont {Ivcheriko},\ and\ \citenamefont
  {Sturman}}]{Belinicher1982a}%
  \BibitemOpen
  \bibfield  {author} {\bibinfo {author} {\bibfnamefont {V.}~\bibnamefont
  {Belinicher}}, \bibinfo {author} {\bibfnamefont {E.~L.}\ \bibnamefont
  {Ivcheriko}}, \ and\ \bibinfo {author} {\bibfnamefont {B.}~\bibnamefont
  {Sturman}},\ }\href@noop {} {\bibfield  {journal} {\bibinfo  {journal} {Zh.
  Eksp. Teor. Fiz.}\ }\textbf {\bibinfo {volume} {83}},\ \bibinfo {pages} {649}
  (\bibinfo {year} {1982})},\ \bibinfo {note} {[J. Exp. Theor. Phys. 56, 359
  (1982)]}\BibitemShut {NoStop}%
\bibitem [{\citenamefont {Sturman}\ and\ \citenamefont
  {Fridkin}(1992)}]{Sturman1992a}%
  \BibitemOpen
  \bibfield  {author} {\bibinfo {author} {\bibfnamefont {B.~I.}\ \bibnamefont
  {Sturman}}\ and\ \bibinfo {author} {\bibfnamefont {V.~M.}\ \bibnamefont
  {Fridkin}},\ }\href {\doibase 10.1201/9780203743416} {\emph {\bibinfo {title}
  {The Photovoltaic and Photorefractive Effects in Noncentrosymmetric protect
  Materials}}}\ (\bibinfo  {publisher} {Gordon and Breach Science Publishers},\
  \bibinfo {address} {Philadelphia},\ \bibinfo {year} {1992})\BibitemShut
  {NoStop}%
\bibitem [{\citenamefont {Sipe}\ and\ \citenamefont
  {Shkrebtii}(2000)}]{Sipe2000a}%
  \BibitemOpen
  \bibfield  {author} {\bibinfo {author} {\bibfnamefont {J.~E.}\ \bibnamefont
  {Sipe}}\ and\ \bibinfo {author} {\bibfnamefont {A.~I.}\ \bibnamefont
  {Shkrebtii}},\ }\bibfield  {title} {\enquote {\bibinfo {title} {Second-order
  optical response in semiconductors},}\ }\href {\doibase
  10.1103/PhysRevB.61.5337} {\bibfield  {journal} {\bibinfo  {journal} {Phys.
  Rev. B}\ }\textbf {\bibinfo {volume} {61}},\ \bibinfo {pages} {5337--5352}
  (\bibinfo {year} {2000})}\BibitemShut {NoStop}%
\bibitem [{\citenamefont {Young}\ and\ \citenamefont
  {Rappe}(2012)}]{Young2012a}%
  \BibitemOpen
  \bibfield  {author} {\bibinfo {author} {\bibfnamefont {Steve~M.}\
  \bibnamefont {Young}}\ and\ \bibinfo {author} {\bibfnamefont {Andrew~M.}\
  \bibnamefont {Rappe}},\ }\bibfield  {title} {\enquote {\bibinfo {title}
  {First principles calculation of the shift current photovoltaic effect in
  ferroelectrics},}\ }\href {\doibase 10.1103/PhysRevLett.109.116601}
  {\bibfield  {journal} {\bibinfo  {journal} {Physical Review Letters}\
  }\textbf {\bibinfo {volume} {109}},\ \bibinfo {pages} {116601} (\bibinfo
  {year} {2012})}\BibitemShut {NoStop}%
\bibitem [{\citenamefont {Cook}\ \emph {et~al.}(2017)\citenamefont {Cook},
  \citenamefont {Fregoso}, \citenamefont {de~Juan}, \citenamefont {Coh},\ and\
  \citenamefont {Moore}}]{Cook2017a}%
  \BibitemOpen
  \bibfield  {author} {\bibinfo {author} {\bibfnamefont {Ashley~M.}\
  \bibnamefont {Cook}}, \bibinfo {author} {\bibfnamefont {B.~M.}\ \bibnamefont
  {Fregoso}}, \bibinfo {author} {\bibfnamefont {F.}~\bibnamefont {de~Juan}},
  \bibinfo {author} {\bibfnamefont {Sinisa}\ \bibnamefont {Coh}}, \ and\
  \bibinfo {author} {\bibfnamefont {J.~E.}\ \bibnamefont {Moore}},\ }\bibfield
  {title} {\enquote {\bibinfo {title} {Design principles for shift current
  photovoltaics},}\ }\href {\doibase 10.1038/ncomms14176} {\bibfield  {journal}
  {\bibinfo  {journal} {Nature Communications}\ }\textbf {\bibinfo {volume}
  {8}},\ \bibinfo {pages} {14176} (\bibinfo {year} {2017})}\BibitemShut
  {NoStop}%
\bibitem [{\citenamefont {Tokura}\ and\ \citenamefont
  {Nagaosa}(2018)}]{Tokura2018a}%
  \BibitemOpen
  \bibfield  {author} {\bibinfo {author} {\bibfnamefont {Y.}~\bibnamefont
  {Tokura}}\ and\ \bibinfo {author} {\bibfnamefont {N.}~\bibnamefont
  {Nagaosa}},\ }\bibfield  {title} {\enquote {\bibinfo {title} {Nonreciprocal
  responses from non-centrosymmetric quantum materials},}\ }\href {\doibase
  10.1038/s41467-018-05759-4} {\bibfield  {journal} {\bibinfo  {journal} {Nat.
  Commun.}\ }\textbf {\bibinfo {volume} {9}},\ \bibinfo {pages} {3740}
  (\bibinfo {year} {2018})}\BibitemShut {NoStop}%
\bibitem [{\citenamefont {Moore}\ and\ \citenamefont
  {Orenstein}(2010)}]{Moore2010a}%
  \BibitemOpen
  \bibfield  {author} {\bibinfo {author} {\bibfnamefont {J.~E.}\ \bibnamefont
  {Moore}}\ and\ \bibinfo {author} {\bibfnamefont {J.}~\bibnamefont
  {Orenstein}},\ }\bibfield  {title} {\enquote {\bibinfo {title}
  {Confinement-induced berry phase and helicity-dependent photocurrents},}\
  }\href {\doibase 10.1103/PhysRevLett.105.026805} {\bibfield  {journal}
  {\bibinfo  {journal} {Physical Review Letters}\ }\textbf {\bibinfo {volume}
  {105}},\ \bibinfo {pages} {026805} (\bibinfo {year} {2010})}\BibitemShut
  {NoStop}%
\bibitem [{\citenamefont {Sodemann}\ and\ \citenamefont
  {Fu}(2015)}]{Sodemann2015a}%
  \BibitemOpen
  \bibfield  {author} {\bibinfo {author} {\bibfnamefont {Inti}\ \bibnamefont
  {Sodemann}}\ and\ \bibinfo {author} {\bibfnamefont {Liang}\ \bibnamefont
  {Fu}},\ }\bibfield  {title} {\enquote {\bibinfo {title} {Quantum nonlinear
  hall effect induced by berry curvature dipole in time-reversal invariant
  materials},}\ }\href {\doibase 10.1103/PhysRevLett.115.216806} {\bibfield
  {journal} {\bibinfo  {journal} {Physical Review Letters}\ }\textbf {\bibinfo
  {volume} {115}},\ \bibinfo {pages} {216806} (\bibinfo {year}
  {2015})}\BibitemShut {NoStop}%
\bibitem [{\citenamefont {Ishizuka}\ \emph {et~al.}(2016)\citenamefont
  {Ishizuka}, \citenamefont {Hayata}, \citenamefont {Ueda},\ and\ \citenamefont
  {Nagaosa}}]{Ishizuka2016a}%
  \BibitemOpen
  \bibfield  {author} {\bibinfo {author} {\bibfnamefont {Hiroaki}\ \bibnamefont
  {Ishizuka}}, \bibinfo {author} {\bibfnamefont {Tomoya}\ \bibnamefont
  {Hayata}}, \bibinfo {author} {\bibfnamefont {Masahito}\ \bibnamefont {Ueda}},
  \ and\ \bibinfo {author} {\bibfnamefont {Naoto}\ \bibnamefont {Nagaosa}},\
  }\bibfield  {title} {\enquote {\bibinfo {title} {Emergent electromagnetic
  induction and adiabatic charge pumping in noncentrosymmetric weyl
  semimetals},}\ }\href {\doibase 10.1103/PhysRevLett.117.216601} {\bibfield
  {journal} {\bibinfo  {journal} {Phys. Rev. Lett.}\ }\textbf {\bibinfo
  {volume} {117}},\ \bibinfo {pages} {216601} (\bibinfo {year}
  {2016})}\BibitemShut {NoStop}%
\bibitem [{\citenamefont {Ishizuka}\ \emph {et~al.}(2017)\citenamefont
  {Ishizuka}, \citenamefont {Hayata}, \citenamefont {Ueda},\ and\ \citenamefont
  {Nagaosa}}]{Ishizuka2017b}%
  \BibitemOpen
  \bibfield  {author} {\bibinfo {author} {\bibfnamefont {Hiroaki}\ \bibnamefont
  {Ishizuka}}, \bibinfo {author} {\bibfnamefont {Tomoya}\ \bibnamefont
  {Hayata}}, \bibinfo {author} {\bibfnamefont {Masahito}\ \bibnamefont {Ueda}},
  \ and\ \bibinfo {author} {\bibfnamefont {Naoto}\ \bibnamefont {Nagaosa}},\
  }\bibfield  {title} {\enquote {\bibinfo {title} {Momentum-space
  electromagnetic induction in weyl semimetals},}\ }\href {\doibase
  10.1103/PhysRevB.95.245211} {\bibfield  {journal} {\bibinfo  {journal} {Phys.
  Rev. B}\ }\textbf {\bibinfo {volume} {95}},\ \bibinfo {pages} {245211}
  (\bibinfo {year} {2017})}\BibitemShut {NoStop}%
\bibitem [{\citenamefont {Morimoto}\ \emph {et~al.}(2018)\citenamefont
  {Morimoto}, \citenamefont {Nakamura}, \citenamefont {Kawasaki},\ and\
  \citenamefont {Nagaosa}}]{Morimoto2018a}%
  \BibitemOpen
  \bibfield  {author} {\bibinfo {author} {\bibfnamefont {Takahiro}\
  \bibnamefont {Morimoto}}, \bibinfo {author} {\bibfnamefont {Masao}\
  \bibnamefont {Nakamura}}, \bibinfo {author} {\bibfnamefont {Masashi}\
  \bibnamefont {Kawasaki}}, \ and\ \bibinfo {author} {\bibfnamefont {Naoto}\
  \bibnamefont {Nagaosa}},\ }\bibfield  {title} {\enquote {\bibinfo {title}
  {Current-voltage characteristic and shot noise of shift current
  photovoltaics},}\ }\href {\doibase 10.1103/PhysRevLett.121.267401} {\bibfield
   {journal} {\bibinfo  {journal} {Phys. Rev. Lett.}\ }\textbf {\bibinfo
  {volume} {121}},\ \bibinfo {pages} {267401} (\bibinfo {year}
  {2018})}\BibitemShut {NoStop}%
\bibitem [{\citenamefont {Ahn}\ \emph {et~al.}(2022)\citenamefont {Ahn},
  \citenamefont {Guo}, \citenamefont {Nagaosa},\ and\ \citenamefont
  {Vishwanath}}]{Ahn2022a}%
  \BibitemOpen
  \bibfield  {author} {\bibinfo {author} {\bibfnamefont {Junyeong}\
  \bibnamefont {Ahn}}, \bibinfo {author} {\bibfnamefont {Guang-Yu}\
  \bibnamefont {Guo}}, \bibinfo {author} {\bibfnamefont {Naoto}\ \bibnamefont
  {Nagaosa}}, \ and\ \bibinfo {author} {\bibfnamefont {Ashvin}\ \bibnamefont
  {Vishwanath}},\ }\bibfield  {title} {\enquote {\bibinfo {title} {Riemannian
  geometry of resonant optical responses},}\ }\href {\doibase
  10.1038/s41567-021-01465-z} {\bibfield  {journal} {\bibinfo  {journal}
  {Nature Physics}\ }\textbf {\bibinfo {volume} {18}},\ \bibinfo {pages}
  {290--295} (\bibinfo {year} {2022})}\BibitemShut {NoStop}%
\bibitem [{\citenamefont {Young}\ \emph {et~al.}(2013)\citenamefont {Young},
  \citenamefont {Zheng},\ and\ \citenamefont {Rappe}}]{Young2013a}%
  \BibitemOpen
  \bibfield  {author} {\bibinfo {author} {\bibfnamefont {Steve~M.}\
  \bibnamefont {Young}}, \bibinfo {author} {\bibfnamefont {Fan}\ \bibnamefont
  {Zheng}}, \ and\ \bibinfo {author} {\bibfnamefont {Andrew~M.}\ \bibnamefont
  {Rappe}},\ }\bibfield  {title} {\enquote {\bibinfo {title} {Prediction of a
  linear spin bulk photovoltaic effect in antiferromagnets},}\ }\href {\doibase
  10.1103/PhysRevLett.110.057201} {\bibfield  {journal} {\bibinfo  {journal}
  {Physical Review Letters}\ }\textbf {\bibinfo {volume} {110}},\ \bibinfo
  {pages} {057201} (\bibinfo {year} {2013})}\BibitemShut {NoStop}%
\bibitem [{\citenamefont {Davydova}\ \emph {et~al.}(2022)\citenamefont
  {Davydova}, \citenamefont {Serbyn},\ and\ \citenamefont
  {Ishizuka}}]{Davydova2022a}%
  \BibitemOpen
  \bibfield  {author} {\bibinfo {author} {\bibfnamefont {Margarita}\
  \bibnamefont {Davydova}}, \bibinfo {author} {\bibfnamefont {Maksym}\
  \bibnamefont {Serbyn}}, \ and\ \bibinfo {author} {\bibfnamefont {Hiroaki}\
  \bibnamefont {Ishizuka}},\ }\bibfield  {title} {\enquote {\bibinfo {title}
  {Symmetry-allowed nonlinear orbital response across the topological phase
  transition in centrosymmetric materials},}\ }\href {\doibase
  10.1103/PhysRevB.105.L121407} {\bibfield  {journal} {\bibinfo  {journal}
  {Physical Review B}\ }\textbf {\bibinfo {volume} {105}},\ \bibinfo {pages}
  {L121407} (\bibinfo {year} {2022})}\BibitemShut {NoStop}%
\bibitem [{\citenamefont {Proskurin}\ \emph {et~al.}(2018)\citenamefont
  {Proskurin}, \citenamefont {Ovchinnikov}, \citenamefont {Kishine},\ and\
  \citenamefont {Stamps}}]{Proskurin2018a}%
  \BibitemOpen
  \bibfield  {author} {\bibinfo {author} {\bibfnamefont {Igor}\ \bibnamefont
  {Proskurin}}, \bibinfo {author} {\bibfnamefont {Alexander~S.}\ \bibnamefont
  {Ovchinnikov}}, \bibinfo {author} {\bibfnamefont {Jun-ichiro}\ \bibnamefont
  {Kishine}}, \ and\ \bibinfo {author} {\bibfnamefont {Robert~L.}\ \bibnamefont
  {Stamps}},\ }\bibfield  {title} {\enquote {\bibinfo {title} {Excitation of
  magnon spin photocurrents in antiferromagnetic insulators},}\ }\href
  {\doibase 10.1103/PhysRevB.98.134422} {\bibfield  {journal} {\bibinfo
  {journal} {Phys. Rev. B}\ }\textbf {\bibinfo {volume} {98}},\ \bibinfo
  {pages} {134422} (\bibinfo {year} {2018})}\BibitemShut {NoStop}%
\bibitem [{\citenamefont {Ishizuka}\ and\ \citenamefont
  {Sato}(2019{\natexlab{a}})}]{Ishizuka2019a}%
  \BibitemOpen
  \bibfield  {author} {\bibinfo {author} {\bibfnamefont {Hiroaki}\ \bibnamefont
  {Ishizuka}}\ and\ \bibinfo {author} {\bibfnamefont {Masahiro}\ \bibnamefont
  {Sato}},\ }\bibfield  {title} {\enquote {\bibinfo {title} {Rectification of
  spin current in inversion-asymmetric magnets with linearly polarized
  electromagnetic waves},}\ }\href {\doibase 10.1103/PhysRevLett.122.197702}
  {\bibfield  {journal} {\bibinfo  {journal} {Phys. Rev. Lett.}\ }\textbf
  {\bibinfo {volume} {122}},\ \bibinfo {pages} {197702} (\bibinfo {year}
  {2019}{\natexlab{a}})}\BibitemShut {NoStop}%
\bibitem [{\citenamefont {Ishizuka}\ and\ \citenamefont
  {Sato}(2022)}]{Ishizuka2022a}%
  \BibitemOpen
  \bibfield  {author} {\bibinfo {author} {\bibfnamefont {Hiroaki}\ \bibnamefont
  {Ishizuka}}\ and\ \bibinfo {author} {\bibfnamefont {Masahiro}\ \bibnamefont
  {Sato}},\ }\bibfield  {title} {\enquote {\bibinfo {title} {Large
  photogalvanic spin current by magnetic resonance in bilayer cr trihalides},}\
  }\href {\doibase 10.1103/PhysRevLett.129.107201} {\bibfield  {journal}
  {\bibinfo  {journal} {Phys. Rev. Lett.}\ }\textbf {\bibinfo {volume} {129}},\
  \bibinfo {pages} {107201} (\bibinfo {year} {2022})}\BibitemShut {NoStop}%
\bibitem [{\citenamefont {Ishizuka}\ and\ \citenamefont
  {Sato}(2019{\natexlab{b}})}]{Ishizuka2019b}%
  \BibitemOpen
  \bibfield  {author} {\bibinfo {author} {\bibfnamefont {Hiroaki}\ \bibnamefont
  {Ishizuka}}\ and\ \bibinfo {author} {\bibfnamefont {Masahiro}\ \bibnamefont
  {Sato}},\ }\bibfield  {title} {\enquote {\bibinfo {title} {Theory for shift
  current of bosons: Photogalvanic spin current in ferrimagnetic and
  antiferromagnetic insulators},}\ }\href {\doibase
  10.1103/PhysRevB.100.224411} {\bibfield  {journal} {\bibinfo  {journal}
  {Phys. Rev. B}\ }\textbf {\bibinfo {volume} {100}},\ \bibinfo {pages}
  {224411} (\bibinfo {year} {2019}{\natexlab{b}})}\BibitemShut {NoStop}%
\bibitem [{\citenamefont {Ishizuka}\ and\ \citenamefont
  {Sato}(2024)}]{Ishizuka2024a}%
  \BibitemOpen
  \bibfield  {author} {\bibinfo {author} {\bibfnamefont {Hiroaki}\ \bibnamefont
  {Ishizuka}}\ and\ \bibinfo {author} {\bibfnamefont {Masahiro}\ \bibnamefont
  {Sato}},\ }\bibfield  {title} {\enquote {\bibinfo {title} {Peltier effect of
  phonons driven by electromagnetic waves},}\ }\href {\doibase
  10.1103/PhysRevB.110.L020303} {\bibfield  {journal} {\bibinfo  {journal}
  {Phys. Rev. B}\ }\textbf {\bibinfo {volume} {110}},\ \bibinfo {pages}
  {L020303} (\bibinfo {year} {2024})}\BibitemShut {NoStop}%
\bibitem [{\citenamefont {Ishizuka}\ and\ \citenamefont
  {Sato}(2025)}]{Ishizuka2025a}%
  \BibitemOpen
  \bibfield  {author} {\bibinfo {author} {\bibfnamefont {Hiroaki}\ \bibnamefont
  {Ishizuka}}\ and\ \bibinfo {author} {\bibfnamefont {Masahiro}\ \bibnamefont
  {Sato}},\ }\bibfield  {title} {\enquote {\bibinfo {title} {Nonlinear optical
  response of truly chiral phonons: Light-induced phonon angular momentum,
  peltier effect, and orbital current},}\ }\href
  {https://arxiv.org/abs/2505.05313} {\bibfield  {journal} {\bibinfo  {journal}
  {arXiv:cond-mat.mes-hall}\ } (\bibinfo {year} {2025})}\BibitemShut {NoStop}%
\bibitem [{\citenamefont {Fujiwara}\ \emph {et~al.}(2023)\citenamefont
  {Fujiwara}, \citenamefont {Kitamura},\ and\ \citenamefont
  {Morimoto}}]{Fujiwara2023a}%
  \BibitemOpen
  \bibfield  {author} {\bibinfo {author} {\bibfnamefont {Kosuke}\ \bibnamefont
  {Fujiwara}}, \bibinfo {author} {\bibfnamefont {Sota}\ \bibnamefont
  {Kitamura}}, \ and\ \bibinfo {author} {\bibfnamefont {Takahiro}\ \bibnamefont
  {Morimoto}},\ }\bibfield  {title} {\enquote {\bibinfo {title} {Nonlinear spin
  current of photoexcited magnons in collinear antiferromagnets},}\ }\href
  {\doibase 10.1103/PhysRevB.107.064403} {\bibfield  {journal} {\bibinfo
  {journal} {Phys. Rev. B}\ }\textbf {\bibinfo {volume} {107}},\ \bibinfo
  {pages} {064403} (\bibinfo {year} {2023})}\BibitemShut {NoStop}%
\bibitem [{\citenamefont {Xu}\ \emph {et~al.}(2021)\citenamefont {Xu},
  \citenamefont {Wang}, \citenamefont {Zhou},\ and\ \citenamefont
  {Li}}]{Xu2021a}%
  \BibitemOpen
  \bibfield  {author} {\bibinfo {author} {\bibfnamefont {Haowei}\ \bibnamefont
  {Xu}}, \bibinfo {author} {\bibfnamefont {Hua}\ \bibnamefont {Wang}}, \bibinfo
  {author} {\bibfnamefont {Jian}\ \bibnamefont {Zhou}}, \ and\ \bibinfo
  {author} {\bibfnamefont {Ju}~\bibnamefont {Li}},\ }\bibfield  {title}
  {\enquote {\bibinfo {title} {Pure spin photocurrent in non‐centrosymmetric
  crystals: bulk spin photovoltaic effect},}\ }\href {\doibase
  10.1038/s41467-021-24541-7} {\bibfield  {journal} {\bibinfo  {journal}
  {Nature Communications}\ }\textbf {\bibinfo {volume} {12}},\ \bibinfo {pages}
  {4330} (\bibinfo {year} {2021})}\BibitemShut {NoStop}%
\bibitem [{\citenamefont {Wang}\ \emph
  {et~al.}(2024{\natexlab{a}})\citenamefont {Wang}, \citenamefont {Meng},
  \citenamefont {Lin}, \citenamefont {Xu}, \citenamefont {Ma}, \citenamefont
  {Kan}, \citenamefont {Chen}, \citenamefont {Huang}, \citenamefont {Chen},
  \citenamefont {Yue}, \citenamefont {Duan}, \citenamefont {Chu},\ and\
  \citenamefont {Sun}}]{Wang2024a}%
  \BibitemOpen
  \bibfield  {author} {\bibinfo {author} {\bibfnamefont {Hongru}\ \bibnamefont
  {Wang}}, \bibinfo {author} {\bibfnamefont {Jing}\ \bibnamefont {Meng}},
  \bibinfo {author} {\bibfnamefont {Jianjun}\ \bibnamefont {Lin}}, \bibinfo
  {author} {\bibfnamefont {Bin}\ \bibnamefont {Xu}}, \bibinfo {author}
  {\bibfnamefont {Hai}\ \bibnamefont {Ma}}, \bibinfo {author} {\bibfnamefont
  {Yucheng}\ \bibnamefont {Kan}}, \bibinfo {author} {\bibfnamefont {Rui}\
  \bibnamefont {Chen}}, \bibinfo {author} {\bibfnamefont {Lujun}\ \bibnamefont
  {Huang}}, \bibinfo {author} {\bibfnamefont {Ye}~\bibnamefont {Chen}},
  \bibinfo {author} {\bibfnamefont {Fangyu}\ \bibnamefont {Yue}}, \bibinfo
  {author} {\bibfnamefont {Chun‐Gang}\ \bibnamefont {Duan}}, \bibinfo
  {author} {\bibfnamefont {Junhao}\ \bibnamefont {Chu}}, \ and\ \bibinfo
  {author} {\bibfnamefont {Lin}\ \bibnamefont {Sun}},\ }\bibfield  {title}
  {\enquote {\bibinfo {title} {Origin of the light‐induced spin currents in
  heavy metal/magnetic insulator bilayers},}\ }\href {\doibase
  10.1038/s41467-024-48710-6} {\bibfield  {journal} {\bibinfo  {journal}
  {Nature Communications}\ }\textbf {\bibinfo {volume} {15}},\ \bibinfo {pages}
  {4362} (\bibinfo {year} {2024}{\natexlab{a}})}\BibitemShut {NoStop}%
\bibitem [{\citenamefont {Huang}\ \emph {et~al.}(2024)\citenamefont {Huang},
  \citenamefont {Yan},\ and\ \citenamefont {Jiang}}]{Huang2024}%
  \BibitemOpen
  \bibfield  {author} {\bibinfo {author} {\bibfnamefont {Zongduo}\ \bibnamefont
  {Huang}}, \bibinfo {author} {\bibfnamefont {Yonghong}\ \bibnamefont {Yan}}, \
  and\ \bibinfo {author} {\bibfnamefont {Feng}\ \bibnamefont {Jiang}},\
  }\bibfield  {title} {\enquote {\bibinfo {title} {Spin current generation in
  an organic antiferromagnet via photo‐excitation},}\ }\href {\doibase
  10.1016/j.orgel.2024.106999} {\bibfield  {journal} {\bibinfo  {journal}
  {Organic Electronics}\ }\textbf {\bibinfo {volume} {126}},\ \bibinfo {pages}
  {106999} (\bibinfo {year} {2024})}\BibitemShut {NoStop}%
\bibitem [{\citenamefont {Zhu}\ \emph {et~al.}(2024)\citenamefont {Zhu},
  \citenamefont {Qu}, \citenamefont {Li}, \citenamefont {Yan},\ and\
  \citenamefont {Liu}}]{Zhu2024}%
  \BibitemOpen
  \bibfield  {author} {\bibinfo {author} {\bibfnamefont {Yudong}\ \bibnamefont
  {Zhu}}, \bibinfo {author} {\bibfnamefont {Junyang}\ \bibnamefont {Qu}},
  \bibinfo {author} {\bibfnamefont {Dan}\ \bibnamefont {Li}}, \bibinfo {author}
  {\bibfnamefont {Yue}\ \bibnamefont {Yan}}, \ and\ \bibinfo {author}
  {\bibfnamefont {Bin}\ \bibnamefont {Liu}},\ }\bibfield  {title} {\enquote
  {\bibinfo {title} {Enhanced photocurrent and spin current in
  two‐dimensional mnncl–mnni lateral heterostructures},}\ }\href {\doibase
  10.1016/j.cplett.2024.141735} {\bibfield  {journal} {\bibinfo  {journal}
  {Chemical Physics Letters}\ }\textbf {\bibinfo {volume} {857}},\ \bibinfo
  {pages} {141735} (\bibinfo {year} {2024})}\BibitemShut {NoStop}%
\bibitem [{\citenamefont {Zhang}\ \emph {et~al.}(2025)\citenamefont {Zhang},
  \citenamefont {Yu}, \citenamefont {Tang}, \citenamefont {Song}, \citenamefont
  {Liu}, \citenamefont {Yang}, \citenamefont {Ge},\ and\ \citenamefont
  {Shen}}]{Zhang2025}%
  \BibitemOpen
  \bibfield  {author} {\bibinfo {author} {\bibfnamefont {Shixiong}\
  \bibnamefont {Zhang}}, \bibinfo {author} {\bibfnamefont {Weizhi}\
  \bibnamefont {Yu}}, \bibinfo {author} {\bibfnamefont {Ning}\ \bibnamefont
  {Tang}}, \bibinfo {author} {\bibfnamefont {Yingming}\ \bibnamefont {Song}},
  \bibinfo {author} {\bibfnamefont {Meifeng}\ \bibnamefont {Liu}}, \bibinfo
  {author} {\bibfnamefont {Xuelin}\ \bibnamefont {Yang}}, \bibinfo {author}
  {\bibfnamefont {Weikun}\ \bibnamefont {Ge}}, \ and\ \bibinfo {author}
  {\bibfnamefont {Bo}~\bibnamefont {Shen}},\ }\bibfield  {title} {\enquote
  {\bibinfo {title} {Efficient generation of spin photocurrent by
  defect‐mediated resonant excitation},}\ }\href {\doibase
  10.1038/s42005-025-02402-9} {\bibfield  {journal} {\bibinfo  {journal}
  {Communications Physics}\ }\textbf {\bibinfo {volume} {8}},\ \bibinfo {pages}
  {493} (\bibinfo {year} {2025})}\BibitemShut {NoStop}%
\bibitem [{\citenamefont {Yang}\ \emph {et~al.}(2025)\citenamefont {Yang},
  \citenamefont {Zhang}, \citenamefont {Zhang}, \citenamefont {Xiao},
  \citenamefont {Jia}, \citenamefont {Chen},\ and\ \citenamefont
  {Zhang}}]{Yang2025}%
  \BibitemOpen
  \bibfield  {author} {\bibinfo {author} {\bibfnamefont {Yaqing}\ \bibnamefont
  {Yang}}, \bibinfo {author} {\bibfnamefont {Zhen}\ \bibnamefont {Zhang}},
  \bibinfo {author} {\bibfnamefont {Liwen}\ \bibnamefont {Zhang}}, \bibinfo
  {author} {\bibfnamefont {Liantuan}\ \bibnamefont {Xiao}}, \bibinfo {author}
  {\bibfnamefont {Suotang}\ \bibnamefont {Jia}}, \bibinfo {author}
  {\bibfnamefont {Jun}\ \bibnamefont {Chen}}, \ and\ \bibinfo {author}
  {\bibfnamefont {Lei}\ \bibnamefont {Zhang}},\ }\href
  {https://arxiv.org/abs/2509.25691} {\enquote {\bibinfo {title}
  {Electric‐field control of pure spin photocurrent in germanene},}\ }
  (\bibinfo {year} {2025}),\ \bibinfo {note} {preprint; available at
  arXiv:2509.25691}\BibitemShut {NoStop}%
\bibitem [{\citenamefont {Movafagh}\ and\ \citenamefont
  {Nikolić}(2025)}]{Movafagh2025}%
  \BibitemOpen
  \bibfield  {author} {\bibinfo {author} {\bibfnamefont {Shahrzad}\
  \bibnamefont {Movafagh}}\ and\ \bibinfo {author} {\bibfnamefont {Predrag}\
  \bibnamefont {Nikolić}},\ }\bibfield  {title} {\enquote {\bibinfo {title}
  {Transverse spin photocurrents in ultrathin topological insulator films},}\
  }\href {\doibase 10.1103/PhysRevB.112.035423} {\bibfield  {journal} {\bibinfo
   {journal} {Physical Review B}\ }\textbf {\bibinfo {volume} {112}},\ \bibinfo
  {pages} {035423} (\bibinfo {year} {2025})}\BibitemShut {NoStop}%
\bibitem [{\citenamefont {Bernevig}\ \emph {et~al.}(2006)\citenamefont
  {Bernevig}, \citenamefont {Hughes},\ and\ \citenamefont
  {Zhang}}]{Bernevig2006a}%
  \BibitemOpen
  \bibfield  {author} {\bibinfo {author} {\bibfnamefont {B.~Andrei}\
  \bibnamefont {Bernevig}}, \bibinfo {author} {\bibfnamefont {Taylor~L.}\
  \bibnamefont {Hughes}}, \ and\ \bibinfo {author} {\bibfnamefont {Shou-Cheng}\
  \bibnamefont {Zhang}},\ }\bibfield  {title} {\enquote {\bibinfo {title}
  {Quantum spin hall effect and topological phase transition in hgte quantum
  wells},}\ }\href {\doibase 10.1126/science.1133734} {\bibfield  {journal}
  {\bibinfo  {journal} {Science}\ }\textbf {\bibinfo {volume} {314}},\ \bibinfo
  {pages} {1757--1761} (\bibinfo {year} {2006})}\BibitemShut {NoStop}%
\bibitem [{\citenamefont {Luttinger}(1956)}]{Luttinger1956}%
  \BibitemOpen
  \bibfield  {author} {\bibinfo {author} {\bibfnamefont {J.~M.}\ \bibnamefont
  {Luttinger}},\ }\bibfield  {title} {\enquote {\bibinfo {title} {The effect of
  a magnetic field on electrons in a periodic potential},}\ }\href
  {https://doi.org/10.1103/PhysRev.102.1030} {\bibfield  {journal} {\bibinfo
  {journal} {Physical Review}\ }\textbf {\bibinfo {volume} {102}},\ \bibinfo
  {pages} {1030--1041} (\bibinfo {year} {1956})}\BibitemShut {NoStop}%
\bibitem [{\citenamefont {Sinova}\ \emph {et~al.}(2015)\citenamefont {Sinova},
  \citenamefont {Valenzuela}, \citenamefont {Wunderlich}, \citenamefont
  {Back},\ and\ \citenamefont {Jungwirth}}]{Sinova2015a}%
  \BibitemOpen
  \bibfield  {author} {\bibinfo {author} {\bibfnamefont {Jairo}\ \bibnamefont
  {Sinova}}, \bibinfo {author} {\bibfnamefont {Sergio~O.}\ \bibnamefont
  {Valenzuela}}, \bibinfo {author} {\bibfnamefont {J.}~\bibnamefont
  {Wunderlich}}, \bibinfo {author} {\bibfnamefont {C.~H.}\ \bibnamefont
  {Back}}, \ and\ \bibinfo {author} {\bibfnamefont {T.}~\bibnamefont
  {Jungwirth}},\ }\bibfield  {title} {\enquote {\bibinfo {title} {Spin hall
  effects},}\ }\href {\doibase 10.1103/RevModPhys.87.1213} {\bibfield
  {journal} {\bibinfo  {journal} {Rev. Mod. Phys.}\ }\textbf {\bibinfo {volume}
  {87}},\ \bibinfo {pages} {1213--1260} (\bibinfo {year} {2015})}\BibitemShut
  {NoStop}%
\bibitem [{\citenamefont {K{\"o}nig}\ \emph {et~al.}(2007)\citenamefont
  {K{\"o}nig}, \citenamefont {Wiedmann}, \citenamefont {Br{\"u}ne},
  \citenamefont {Roth}, \citenamefont {Buhmann}, \citenamefont {Molenkamp},
  \citenamefont {Qi},\ and\ \citenamefont {Zhang}}]{Konig2007a}%
  \BibitemOpen
  \bibfield  {author} {\bibinfo {author} {\bibfnamefont {Markus}\ \bibnamefont
  {K{\"o}nig}}, \bibinfo {author} {\bibfnamefont {Steffen}\ \bibnamefont
  {Wiedmann}}, \bibinfo {author} {\bibfnamefont {Christoph}\ \bibnamefont
  {Br{\"u}ne}}, \bibinfo {author} {\bibfnamefont {Andreas}\ \bibnamefont
  {Roth}}, \bibinfo {author} {\bibfnamefont {Hartmut}\ \bibnamefont {Buhmann}},
  \bibinfo {author} {\bibfnamefont {Laurens~W.}\ \bibnamefont {Molenkamp}},
  \bibinfo {author} {\bibfnamefont {Xiao-Liang}\ \bibnamefont {Qi}}, \ and\
  \bibinfo {author} {\bibfnamefont {Shou-Cheng}\ \bibnamefont {Zhang}},\
  }\bibfield  {title} {\enquote {\bibinfo {title} {Quantum spin hall insulator
  state in hgte quantum wells},}\ }\href {\doibase 10.1126/science.1148047}
  {\bibfield  {journal} {\bibinfo  {journal} {Science}\ }\textbf {\bibinfo
  {volume} {318}},\ \bibinfo {pages} {766--770} (\bibinfo {year}
  {2007})}\BibitemShut {NoStop}%
\bibitem [{\citenamefont {Bernevig}\ \emph {et~al.}(2005)\citenamefont
  {Bernevig}, \citenamefont {Hughes},\ and\ \citenamefont
  {Zhang}}]{bernevig2005a}%
  \BibitemOpen
  \bibfield  {author} {\bibinfo {author} {\bibfnamefont {B.~Andrei}\
  \bibnamefont {Bernevig}}, \bibinfo {author} {\bibfnamefont {Taylor~L.}\
  \bibnamefont {Hughes}}, \ and\ \bibinfo {author} {\bibfnamefont {Shou-Cheng}\
  \bibnamefont {Zhang}},\ }\bibfield  {title} {\enquote {\bibinfo {title}
  {Orbitronics: {The} {Intrinsic} {Orbital} {Current} in \$p\$-{Doped}
  {Silicon}},}\ }\href {\doibase 10.1103/PhysRevLett.95.066601} {\bibfield
  {journal} {\bibinfo  {journal} {Physical Review Letters}\ }\textbf {\bibinfo
  {volume} {95}},\ \bibinfo {pages} {066601} (\bibinfo {year}
  {2005})}\BibitemShut {NoStop}%
\bibitem [{\citenamefont {Wang}\ \emph
  {et~al.}(2024{\natexlab{b}})\citenamefont {Wang}, \citenamefont {Chen},
  \citenamefont {Yang}, \citenamefont {Hu}, \citenamefont {Li}, \citenamefont
  {Wang}, \citenamefont {Zhang},\ and\ \citenamefont {Jiang}}]{Wang2024}%
  \BibitemOpen
  \bibfield  {author} {\bibinfo {author} {\bibfnamefont {Ping}\ \bibnamefont
  {Wang}}, \bibinfo {author} {\bibfnamefont {Feng}\ \bibnamefont {Chen}},
  \bibinfo {author} {\bibfnamefont {Yuhe}\ \bibnamefont {Yang}}, \bibinfo
  {author} {\bibfnamefont {Shuai}\ \bibnamefont {Hu}}, \bibinfo {author}
  {\bibfnamefont {Yue}\ \bibnamefont {Li}}, \bibinfo {author} {\bibfnamefont
  {Wenhong}\ \bibnamefont {Wang}}, \bibinfo {author} {\bibfnamefont {Delin}\
  \bibnamefont {Zhang}}, \ and\ \bibinfo {author} {\bibfnamefont {Yong}\
  \bibnamefont {Jiang}},\ }\bibfield  {title} {\enquote {\bibinfo {title}
  {Orbitronics: Mechanisms, materials and devices},}\ }\href {\doibase
  10.1002/aelm.202400554} {\bibfield  {journal} {\bibinfo  {journal} {Advanced
  Electronic Materials}\ }\textbf {\bibinfo {volume} {11}} (\bibinfo {year}
  {2024}{\natexlab{b}}),\ 10.1002/aelm.202400554}\BibitemShut {NoStop}%
\bibitem [{\citenamefont {Moon}\ \emph {et~al.}(2013)\citenamefont {Moon},
  \citenamefont {Xu}, \citenamefont {Kim},\ and\ \citenamefont
  {Balents}}]{Moon2013a}%
  \BibitemOpen
  \bibfield  {author} {\bibinfo {author} {\bibfnamefont {Eun-Gook}\
  \bibnamefont {Moon}}, \bibinfo {author} {\bibfnamefont {Cenke}\ \bibnamefont
  {Xu}}, \bibinfo {author} {\bibfnamefont {Yong~Baek}\ \bibnamefont {Kim}}, \
  and\ \bibinfo {author} {\bibfnamefont {Leon}\ \bibnamefont {Balents}},\
  }\bibfield  {title} {\enquote {\bibinfo {title} {Non-fermi-liquid and
  topological states with strong spin-orbit coupling},}\ }\href {\doibase
  10.1103/PhysRevLett.111.206401} {\bibfield  {journal} {\bibinfo  {journal}
  {Phys. Rev. Lett.}\ }\textbf {\bibinfo {volume} {111}},\ \bibinfo {pages}
  {206401} (\bibinfo {year} {2013})}\BibitemShut {NoStop}%
\bibitem [{\citenamefont {Terasawa}\ and\ \citenamefont
  {Ishizuka}(2024)}]{Terasawa2024a}%
  \BibitemOpen
  \bibfield  {author} {\bibinfo {author} {\bibfnamefont {Ryunosuke}\
  \bibnamefont {Terasawa}}\ and\ \bibinfo {author} {\bibfnamefont {Hiroaki}\
  \bibnamefont {Ishizuka}},\ }\bibfield  {title} {\enquote {\bibinfo {title}
  {Anomalous hall effect by chiral spin textures in the two-dimensional
  luttinger model},}\ }\href {\doibase 10.1103/PhysRevB.109.L060407} {\bibfield
   {journal} {\bibinfo  {journal} {Phys. Rev. B}\ }\textbf {\bibinfo {volume}
  {109}},\ \bibinfo {pages} {L060407} (\bibinfo {year} {2024})}\BibitemShut
  {NoStop}%
\bibitem [{\citenamefont {Zhang}\ \emph {et~al.}(2018)\citenamefont {Zhang},
  \citenamefont {Wang}, \citenamefont {Ruan}, \citenamefont {Yao},\ and\
  \citenamefont {Zhang}}]{Zhang2018}%
  \BibitemOpen
  \bibfield  {author} {\bibinfo {author} {\bibfnamefont {Dongqin}\ \bibnamefont
  {Zhang}}, \bibinfo {author} {\bibfnamefont {Huaiqiang}\ \bibnamefont {Wang}},
  \bibinfo {author} {\bibfnamefont {Jiawei}\ \bibnamefont {Ruan}}, \bibinfo
  {author} {\bibfnamefont {Ge}~\bibnamefont {Yao}}, \ and\ \bibinfo {author}
  {\bibfnamefont {Haijun}\ \bibnamefont {Zhang}},\ }\bibfield  {title}
  {\enquote {\bibinfo {title} {Engineering topological phases in the luttinger
  semimetal $\alpha$‑sn},}\ }\href {\doibase 10.1103/PhysRevB.97.195139} {\bibfield
   {journal} {\bibinfo  {journal} {Physical Review B}\ }\textbf {\bibinfo
  {volume} {97}},\ \bibinfo {pages} {195139} (\bibinfo {year}
  {2018})}\BibitemShut {NoStop}%
\bibitem [{\citenamefont {Anh}\ \emph {et~al.}(2021)\citenamefont {Anh},
  \citenamefont {Takase}, \citenamefont {Chiba}, \citenamefont {Kota},
  \citenamefont {Takiguchi},\ and\ \citenamefont {Tanaka}}]{Anh2021a}%
  \BibitemOpen
  \bibfield  {author} {\bibinfo {author} {\bibfnamefont {Le~Duc}\ \bibnamefont
  {Anh}}, \bibinfo {author} {\bibfnamefont {Kengo}\ \bibnamefont {Takase}},
  \bibinfo {author} {\bibfnamefont {Takahiro}\ \bibnamefont {Chiba}}, \bibinfo
  {author} {\bibfnamefont {Yohei}\ \bibnamefont {Kota}}, \bibinfo {author}
  {\bibfnamefont {Kosuke}\ \bibnamefont {Takiguchi}}, \ and\ \bibinfo {author}
  {\bibfnamefont {Masaaki}\ \bibnamefont {Tanaka}},\ }\bibfield  {title}
  {\enquote {\bibinfo {title} {Elemental topological dirac semimetal
  {$\alpha$}-{Sn} with high quantum mobility},}\ }\href {\doibase
  10.1002/adma.202104645} {\bibfield  {journal} {\bibinfo  {journal} {Advanced
  Materials}\ }\textbf {\bibinfo {volume} {33}},\ \bibinfo {pages} {2104645}
  (\bibinfo {year} {2021})}\BibitemShut {NoStop}%
\bibitem [{\citenamefont {Duan}(2022)}]{Duan2022a}%
  \BibitemOpen
  \bibfield  {author} {\bibinfo {author} {\bibfnamefont {Jinsong}\ \bibnamefont
  {Duan}},\ }\href {https://arxiv.org/abs/2205.08686} {\enquote {\bibinfo
  {title} {Dielectric constant of gray tin: A first-principles study},}\ }
  (\bibinfo {year} {2022}),\ \Eprint {http://arxiv.org/abs/2205.08686}
  {arXiv:2205.08686 [cond-mat.mtrl-sci]} \BibitemShut {NoStop}%
\bibitem [{\citenamefont {Maekawa}\ \emph {et~al.}(2012)\citenamefont
  {Maekawa}, \citenamefont {Valenzuela}, \citenamefont {Saitoh},\ and\
  \citenamefont {Kimura}}]{Maekawa2012a}%
  \BibitemOpen
  \bibinfo {editor} {\bibfnamefont {S.}~\bibnamefont {Maekawa}}, \bibinfo
  {editor} {\bibfnamefont {S.~O.}\ \bibnamefont {Valenzuela}}, \bibinfo
  {editor} {\bibfnamefont {E.}~\bibnamefont {Saitoh}}, \ and\ \bibinfo {editor}
  {\bibfnamefont {T.}~\bibnamefont {Kimura}},\ eds.,\ \href
  {https://academic.oup.com/book/10046} {\emph {\bibinfo {title} {Spin
  Current}}}\ (\bibinfo  {publisher} {Oxford University Press},\ \bibinfo
  {address} {Oxford, England},\ \bibinfo {year} {2012})\BibitemShut {NoStop}%
\bibitem [{\citenamefont {Dong}\ \emph {et~al.}(2025)\citenamefont {Dong},
  \citenamefont {Cao}, \citenamefont {Tan},\ and\ \citenamefont
  {Fei}}]{Dong2025}%
  \BibitemOpen
  \bibfield  {author} {\bibinfo {author} {\bibfnamefont {Ruizhi}\ \bibnamefont
  {Dong}}, \bibinfo {author} {\bibfnamefont {Ranquan}\ \bibnamefont {Cao}},
  \bibinfo {author} {\bibfnamefont {Dian}\ \bibnamefont {Tan}}, \ and\ \bibinfo
  {author} {\bibfnamefont {Ruixiang}\ \bibnamefont {Fei}},\ }\bibfield  {title}
  {\enquote {\bibinfo {title} {Crystal symmetry selected pure spin photocurrent
  in altermagnetic insulators},}\ }\href {\doibase 10.1103/PhysRevB.111.195210}
  {\bibfield  {journal} {\bibinfo  {journal} {Physical Review B}\ }\textbf
  {\bibinfo {volume} {111}},\ \bibinfo {pages} {195210} (\bibinfo {year}
  {2025})}\BibitemShut {NoStop}%
\end{thebibliography}
%

\end{document}